\documentclass{aa} 
\usepackage{txfonts, newtxtext,newtxmath, bm, amsmath, amssymb, booktabs, array, enumerate, multirow, booktabs, threeparttable, soul, color, graphicx, extarrows} 
\usepackage[normalem]{ulem}
\usepackage[T1]{fontenc}
\usepackage[colorlinks=true, linkcolor=blue, citecolor=blue, urlcolor=blue]{hyperref} 
\usepackage{orcidlink}
\usepackage[switch]{lineno}

\newcommand{\DOA}{Department of Astronomy, School of Physics, Peking University, Beijing 100871, China}
\newcommand{\KIAA}{Kavli Institute for Astronomy and Astrophysics, Peking University, Beijing 100871, China}
\newcommand{\GX}{Guangxi Key Laboratory for Relativistic Astrophysics, School of Physical Science and Technology, Guangxi University, Nanning 530004, China}
\newcommand{\NAOC}{National Astronomical Observatories, Chinese Academy of Sciences, Beijing 100101, China}

\begin{document} 

\title{Novae breves from magnetar giant flares: Potential probes of neutron star crusts}

\titlerunning{Novae breves as probes of NS crusts}
\authorrunning{Zhong et al.}


\author{Jiahang Zhong\inst{1, 2}\orcidlink{0009-0008-2673-1764}
		\and
		Qiu-Hong Chen\inst{3}\orcidlink{0009-0006-8625-5283}
		\and
		Yacheng Kang\inst{1, 2,\thanks{email: yckang@stu.pku.edu.cn}}\orcidlink{0000-0001-7402-4927}
		\and
		Hong-Bo Li\inst{2}\orcidlink{0000-0002-4850-8351}
		\and
		Jinghao Zhang\inst{1, 2}\orcidlink{0009-0002-1101-2798}
		\and\\
		Meng-Hua Chen\inst{3}\orcidlink{0000-0001-8406-8683}
		\and
		Lĳing Shao\inst{2, 4,\thanks{email: lshao@pku.edu.cn}}\orcidlink{0000-0002-1334-8853}}
		

\institute{\DOA \and \KIAA \and \GX \and \NAOC}


\date{Received June 1, 2026 / Accepted June 1, 2026}


\abstract
{Matter ejected from the magnetar crust during giant flares (GFs) may undergo $r$-process nucleosynthesis, producing short-lived optical transients termed ``novae breves''. Although intrinsically much fainter than kilonovae from compact binary mergers, novae breves may occur within or near the Galaxy, making them promising observational targets.}
{We aim to investigate how the neutron star (NS) equation of state (EOS) and the mass of the central magnetar affect the ejecta properties following GFs and the resulting nova brevis emission.}
{We employ a semi-analytical ejecta model combined with nuclear reaction network calculations to compute nucleosynthesis yields and multi-band light curves for different EOSs and magnetar masses, and assess their detectability with current and future facilities.}
{We find that variations in the EOS and magnetar mass modify the ejecta mass and its density and velocity distributions, etc., leading to observable differences in nova brevis light curves. In particular, both the peak luminosity and the characteristic peak timescale are EOS-dependent. Assuming a fixed Galactic magnetar mass of $1.4\,\mathrm{M}_\odot$ and taking the $u$ band as an example, we find that the minimum apparent AB magnitudes range from $\sim 7\,\mathrm{mag}$ (H4 EOS) to $\sim 8.5\,\mathrm{mag}$ (WFF EOS) with peak timescales of $\simeq 10^{2}$--$10^{3}\,\mathrm{s}$. A more massive magnetar produces fainter emission with a shorter peak timescale. For a magnetar mass of $1.4\,\mathrm{M}_\odot,$ novae breves associated with known magnetars may reach peak luminosities of $\sim 10^{37}$--$10^{39}\,\mathrm{erg} \, \mathrm{s}^{-1}$, enabling targeted searches, particularly following high-energy GF alerts. Larger ejecta masses yield higher peak luminosities. Moreover, a detection horizon of ${\simeq 10\,\mathrm{Mpc}}$ or beyond is achievable with current and future facilities, allowing searches for novae breves from previously unknown magnetars in the Local Volume.}
{Although challenging, detection of such rapidly evolving transients is feasible. Future searches for novae breves can help establish their observational existence and improve our understanding of the NS EOS and crustal properties.}


          
\maketitle


\section{Introduction}
\label{ sec:intro }

The Universe hosts hundreds of chemical elements and a vast number of isotopes, yet explaining their cosmic abundance patterns has long remained a central problem in nuclear astrophysics. Early theoretical efforts include the seminal Big Bang nucleosynthesis (BBN) framework proposed by \citet{Gamow:1948pob} and \citet{Alpher:1948ve}, which aimed to account for the origin of the chemical elements in the early Universe. It is now well established, however, that BBN can produce only the lightest nuclei, primarily hydrogen, helium, and their isotopes \citep{Arcones:2022jer}. Consequently, the synthesis of heavier elements must proceed through astrophysical processes occurring at later cosmic times.

A major breakthrough was achieved by the B$^2$FH theory \citep{Burbidge:1957vc}, which demonstrated that the stellar nucleosynthesis, together with mass loss and explosive events such as supernovae (SNe), plays a dominant role in shaping the chemical enrichment of the Universe over cosmic time. Within this framework, the rapid neutron-capture ($r$-)process is particularly crucial for producing elements heavier than iron.\footnote{We refer readers to \citet{Arcones:2022jer} for a more comprehensive review.} The $r$-process operates under extremely neutron-rich conditions, in which neutron captures proceed on timescales much shorter than those of $\beta$-decay, driving nuclei toward very high mass numbers before they subsequently decay back toward stability and populate the heavy-element abundance pattern.

However, an increasing number of studies have suggested that achieving sufficiently neutron-rich conditions within stellar interiors or proto-neutron-star (proto-NS) winds following SNe is challenging, typically allowing the synthesis of nuclei only up to mass numbers $\mathrm{A} \lesssim 110$ \citep[see e.g.,][]{Wanajo:2010ig, Martinez-Pinedo:2012eaj, Roberts:2012um}. These limitations have motivated exploration of alternative astrophysical sites for $r$-process nucleosynthesis, with compact binary mergers involving NSs, such as NS--NS and NS--black hole (NS--BH) mergers, emerging as leading candidates \citep{Thielemann:2017acv, Chen:2024acv}. Such mergers can eject substantial amounts of neutron-rich material at high velocities through tidal disruption, shock heating, and post-merger winds, thereby providing favorable conditions for $r$-process nucleosynthesis. Moreover, compact binary mergers are expected to produce a variety of observable electromagnetic (EM) transients \citep[see e.g.,][]{Nakar:2019fza}. Among these, kilonova emissions were initially proposed to arise from the radioactive decay of freshly synthesized $r$-process nuclei in the merger ejecta, radiating primarily in the ultraviolet (UV), optical, and infrared (IR) bands \citep{Li:1998bw, Metzger:2010sy}. As such, kilonovae are regarded as a distinctive probe of compact binary mergers \citep{Fernandez:2015use, Barnes:2016umi, Metzger:2019zeh}. Notably, this picture was confirmed by the co-detection of the kilonova AT2017gfo \citep{Andreoni:2017ppd, Arcavi:2017xiz, Chornock:2017sdf, Coulter:2017wya, Cowperthwaite:2017dyu, TOROS:2017pqe, Drout:2017ijr, Evans:2017mmy, Hu:2017tlb, Kasliwal:2017ngb, Kilpatrick:2017mhz, Lipunov:2017dwd, McCully:2017lgx, Nicholl:2017ahq, Pian:2017gtc, Shappee:2017zly, Smartt:2017fuw, DES:2017kbs, Tanvir:2017pws, J-GEM:2017tyx, Valenti:2017ngx} in association with the first multi-messenger gravitational-wave event GW170817 from a NS--NS merger \citep{LIGOScientific:2017vwq, LIGOScientific:2017ync, Margutti:2020xbo}. 

Despite these theoretical and observational successes, growing evidence suggests that compact binary mergers alone may not fully account for the total observed abundances of heavy elements in the Galaxy \citep{Cote:2018qku, Zevin:2019obe, vandeVoort:2019bxg, Thielemann:2020qmv, 2021ApJ...920L..32T}, nor in extremely metal-poor stars \citep{Qian:2007vq, Sneden:2008pbv, 2013AJ....145...26R, 2024ApJ...974..232O}. These discrepancies indicate that additional $r$-process sources must contribute to the galactic chemical evolution. Among the proposed candidates, magnetar giant flares (MGFs), originating from magnetars with ultra-strong magnetic fields of $10^{14}$--$10^{15}\,\mathrm{G}$ \citep{Duncan:1992hi}, have recently drawn attention as promising sites for ejecting small amounts of neutron-rich material capable of undergoing $r$-process nucleosynthesis \citep{Cehula:2023hdh, Patel:2025frn, Patel:2025tse, Patel:2026ynu}. Sudden dissipation of ultrastrong magnetic fields above magnetars can trigger a variety of high-energy transient outbursts, among which the most violent are MGFs \citep{Turolla:2015mwa, popov:2007uv}. A generic mechanism for such localized magnetic energy dissipation is magnetic reconnection, which breaks and subsequently reconfigures magnetic field lines. During a MGF, the enhanced external magnetic pressure may drive shocks into the magnetar crust, leading to dissociation of baryonic matter. Before the magnetic field lines fully reconnect, a small fraction of the crust can subsequently be expelled \citep{Demidov:2022jrv, Cehula:2023hdh}. The resulting ejecta are expected to be neutron-rich, thereby providing favorable conditions for $r$-process nucleosynthesis and its associated follow-up EM transients. 

On the one hand, since the total ejecta mass from a single magnetar is much smaller than that produced in compact binary mergers, the associated optical transients should have lower peak luminosities and shorter characteristic timescales than typical merger-driven kilonovae. Accordingly, such theoretically predicted thermal optical transients following MGFs are referred to as ``novae breves'' \citep{Cehula:2023hdh, Patel:2025frn, Patel:2025tse}. On the other hand, because these events can occur within our Galaxy, their proximity may significantly enhance their detectability. Indeed, observational evidence for $r$-process nucleosynthesis has already been reported in the MGF of SGR 1806$-$20 \citep{Patel:2025tse}. If more events are detected in the future, they would not only help improve our understanding of the contribution of different astrophysical sources to the galactic heavy-element inventory, but also provide a unique opportunity to probe magnetar physics, particularly the properties of NS crusts.

Motivated by this, in this work we aim to demonstrate that novae breves associated with MGFs can serve as potential probes of NS crust physics. Specifically, we extend \citet{Patel:2025frn, Patel:2025tse} by investigating how different NS equations of state (EOSs) and magnetar masses affect the properties of the ejecta launched following MGFs and, consequently, the resulting nova brevis emission. We show that variations in the EOS and magnetar mass can alter the ejecta properties, such as the total ejetca masses and its density and velocity distributions, leading to observable signatures in the multi-band light curves of novae breves. In particular, both the peak luminosity and the characteristic peak timescale are EOS-dependent. Assuming a fixed Galactic magnetar mass of $1.4\,\mathrm{M}_\odot$ and taking the $u$ band as an example, we find that the minimum apparent AB magnitudes (i.e., at peak brightness) range from $\sim 7\,\mathrm{mag}$ for the stiff H4 EOS to $\sim 8.5\,\mathrm{mag}$ for the soft WFF EOS, with peak timescales of $\simeq 10^{2}$--$10^{3}\,\mathrm{s}$. A more massive central magnetar generally results in a lower peak luminosity and a shorter peak timescale. Consequently, the detection of such rapidly evolving transients is challenging, but not entirely prohibitive, and a positive detection may offer an indirect avenue for constraining the magnetar crustal properties. We further investigate the detection prospects of these fast and faint novae breves with current and next-generation ground-based and space-borne telescopes. Although not every MGF produces an observable nova brevis emission, targeted monitoring offers a promising pathway toward realistic detections from known magnetars, which may generate novae breves with peak bolometric luminosities of $\sim 10^{37}$--$10^{39}\,\mathrm{erg} \, \mathrm{s}^{-1}$ for a magnetar mass of $1.4\,\mathrm{M}_\odot$. For unknown magnetars, instruments with rapid slewing capabilities and wide fields of view provide a complementary advantage in capturing such short-lived nova brevis transients. A detection horizon of ${\simeq 10\,\mathrm{Mpc}}$, or even beyond, achievable with current and future facilities, would encompass the nearest star-forming galaxies within the Local Volume. Overall, our results indicate that future searches for novae breves are feasible. They can not only establish their observational existence, but also improve our understanding of the NS EOS and crustal properties.
 
The structure of this paper is as follows. In Sect.~\ref{ sec:ejecta }, we briefly present the NS EOSs used in this work, together with the resulting properties of the unbound ejecta and subsequent nucleosynthesis. In Sect.~\ref{ sec:result }, we present the multi-band light curves of novae breves for different EOSs and magnetar masses, and discuss the detection prospects of these short-lived transients with current and next-generation facilities. Finally, we summarize in Sect.~\ref{ sec:conclu }.


\section{Baryonic ejecta after MGFs}
\label{ sec:ejecta }


\subsection{NS EOSs}
\label{ sec:crust_eos }

\begin{figure}[t]
    \resizebox{\hsize}{!}{\includegraphics{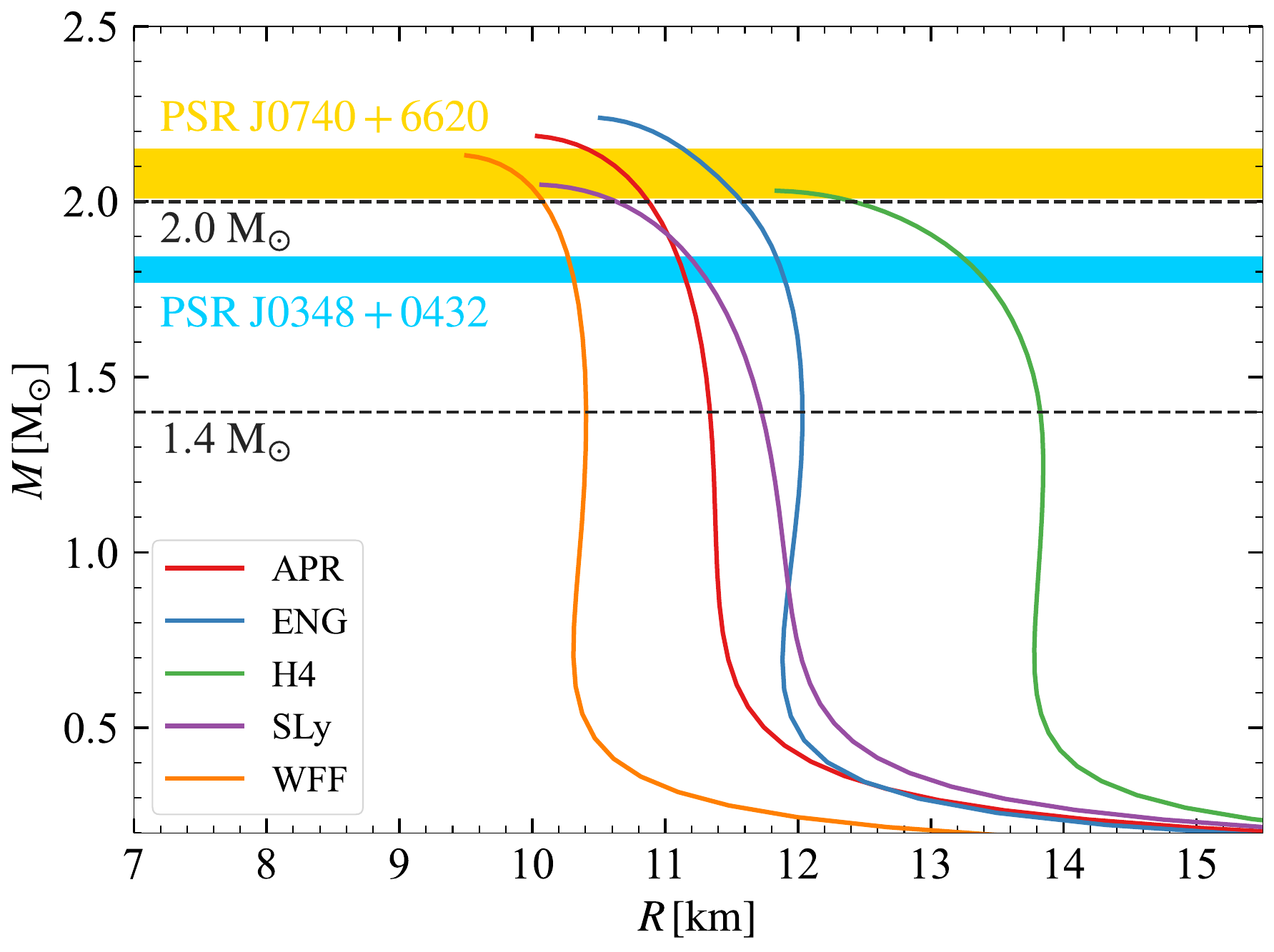}}
    \caption{$M$--$R$ relations for five representative NS EOSs: APR, ENG, H4, SLy, and WFF \citep{Lattimer:2000nx, Ozel:2016oaf, Lattimer:2021emm}. Different colors correspond to different EOSs. In this work, we restrict our analysis to two magnetar masses, $M = 1.4\,\mathrm{M}_\odot$ and $M = 2.0\,\mathrm{M}_\odot$, indicated by the black horizontal dashed lines. Observational constraints from PSR J0740+6620 \citep[yellow band;][]{Fonseca:2021wxt} and PSR J0348+0432 \citep[blue band;][]{Saffer:2024tlb} are shown in the figure.}
\label{ fig:MR relation }
\end{figure}

Before investigating the properties of the unbound ejecta and subsequent nucleosynthesis, we first specify the NS EOSs adopted in this work. To explore as broadly as possible the allowed EOS parameter space and the resulting diversity in nova brevis emissions, we consider five representative EOSs: APR, ENG, H4, SLy, and WFF \citep[see e.g.,][]{Lattimer:2000nx, Ozel:2016oaf, Lattimer:2021emm, Gao:2021uus}. These EOSs differ in their particle contents \citep[e.g., the H4 EOS includes hyperons;][]{Lackey:2005tk}, model constructions \citep[e.g., the non-relativistic Skyrme-Lyon model for the SLy EOS;][]{CHABANAT:1998}, and underlying theoretical approaches \citep[e.g., variational methods for the APR and WFF EOSs, and Dirac-Brueckner Hartree-Fock calculations for the ENG EOS;][]{Akmal:1998cf, Engvik:1995gn, Wiringa:1988}. The detailed compositions of these EOSs are summarized in Table~1 of \citet{Lattimer:2000nx}. The corresponding NS mass--radius ($M$--$R$) relations are shown in Fig.~\ref{ fig:MR relation }. For a given EOS and magnetar mass $M$, the stellar radius $R$, as well as the internal pressure and density profiles, can be self-consistently determined via solving the Tolman-Oppenheimer-Volkoff equations. In this work, we restrict our analysis to two representative magnetar masses, $M = 1.4\,\mathrm{M}_\odot$ and $M = 2.0\,\mathrm{M}_\odot$, and examine the resulting differences in the properties of the corresponding novae breves. Since our study focuses on the outer crust of magnetars, we retain only the region with densities ${\rho \lesssim 4 \times 10^{11} \,\mathrm{g\,cm^{-3}}}$ for further analysis. This choice is well motivated, as only unbound crustal material contributes to the ejecta and, consequently, to the nova brevis emission.

\subsection{Unbound ejecta properties}
\label{ sec:ejecta_prop }

To obtain the properties of the unbound ejecta for different EOSs and magnetar masses, we follow the model framework introduced by \citet{Patel:2025frn, Patel:2025tse}, which is briefly summarized in Appendix~\ref{ app:A }. The resulting unbound ejecta masses are listed in Table~\ref{ tab:ejecta }. As shown in Table~\ref{ tab:ejecta }, for a given magnetar mass, stiffer EOSs generally yield more massive ejecta. This trend naturally arises from the larger stellar radii and weaker gravitational binding associated with stiffer EOSs, which facilitate mass ejection after MGFs. In addition to the total ejecta mass, other EOS-dependent properties, such as the density and velocity distributions, are discussed in Appdix~\ref{ app:A }.


\subsection{Nucleosynthesis for novae breves}
\label{ sec:nucleosynthesis }

Following \citet{Patel:2025frn}, we divide the post-shock unbound ejecta into 30 concentric layers, evolve each layer independently, and track their dynamical and thermal evolution together with the associated nucleosynthesis. For a given choice of EOS and magnetar mass, we employ the nuclear reaction network, \texttt{SkyNet}\footnote{\url{https://bitbucket.org/jlippuner/skynet/src/master/}} \citep{Lippuner:2015gwa, Lippuner:2017tyn}, to compute the time-dependent nucleosynthesis and total elemental yields of the ejecta. Based on these nucleosynthesis yields, we subsequently compute the light curves of novae breves, explicitly accounting for the contribution from $r$-process nuclei. Additional nucleosynthesis results are provided in Appdix~\ref{ app:B }. Details of the light-curve modeling are also described, including several modifications relative to \citet{Patel:2025frn, Patel:2025tse}, particularly in the treatment of radioactive heating rates. For the decay energy of heavy $r$-process nuclei, we adopt the latest data from the Evaluated Nuclear Data File library \citep[ENDF/B-VIII.1;][]{Nobre:2025xlg}. Our results show that the total nucleosynthesis yields of the ejecta exhibit non-negligible variations across different EOSs  and magnetar masses, especially for heavy elements with $\mathrm{A} \gtrsim 140$ (see Fig.~\ref{ fig:abundance }). These EOS-dependent nucleosynthesis signatures are expected to influence the subsequent nova brevis emission and, consequently, their detectability.

\begin{table}[t]
	\caption{Magnetar radius $R$ and unbound ejecta mass $M_{\mathrm{ej}}$ for different EOSs at fixed magnetar masses $M$.}
    \renewcommand{\arraystretch}{2}
    \centering
    \setlength{\tabcolsep}{0.6cm}
    {\begin{tabular}{c c c}
    \toprule 
    \toprule\\
    [-2.5em]
    EOS    & $R\ [\mathrm{km}]$    & $M_{\mathrm{ej}}\  [\times\,10^{-8}\,\mathrm{M}_{\odot}]$ \\
    [0.3em]
    \cline{1-3}\\
    [-2em]
    APR    & 11.34 (10.87)    & 10.6 (3.78)\\
    ENG    & 12.03 (11.58)    & 14.3 (5.09)\\
    H4    & 13.83 (12.42)    & 30.1 (7.70)\\
    SLy    & 11.73 (10.63)    & 12.7 (3.35)\\
    WFF    & 10.41 (10.08)    & 6.58 (2.44)\\
    \bottomrule
    \end{tabular}}
    \tablefoot{For each EOS, results are reported for $M = 1.4\,\mathrm{M}_\odot$, and for $M = 2.0\,\mathrm{M}_\odot$ in parentheses. The magnetic field strength is taken to be $B \simeq 10^{15}\,\mathrm{G}$.}
\label{ tab:ejecta }
\end{table}

\section{Result} 
\label{ sec:result }


\subsection{Light curves of novae breves}
\label{ sec:nova_brevis }

\begin{figure}[t]
    \centering
    \resizebox{\hsize}{!}{\includegraphics{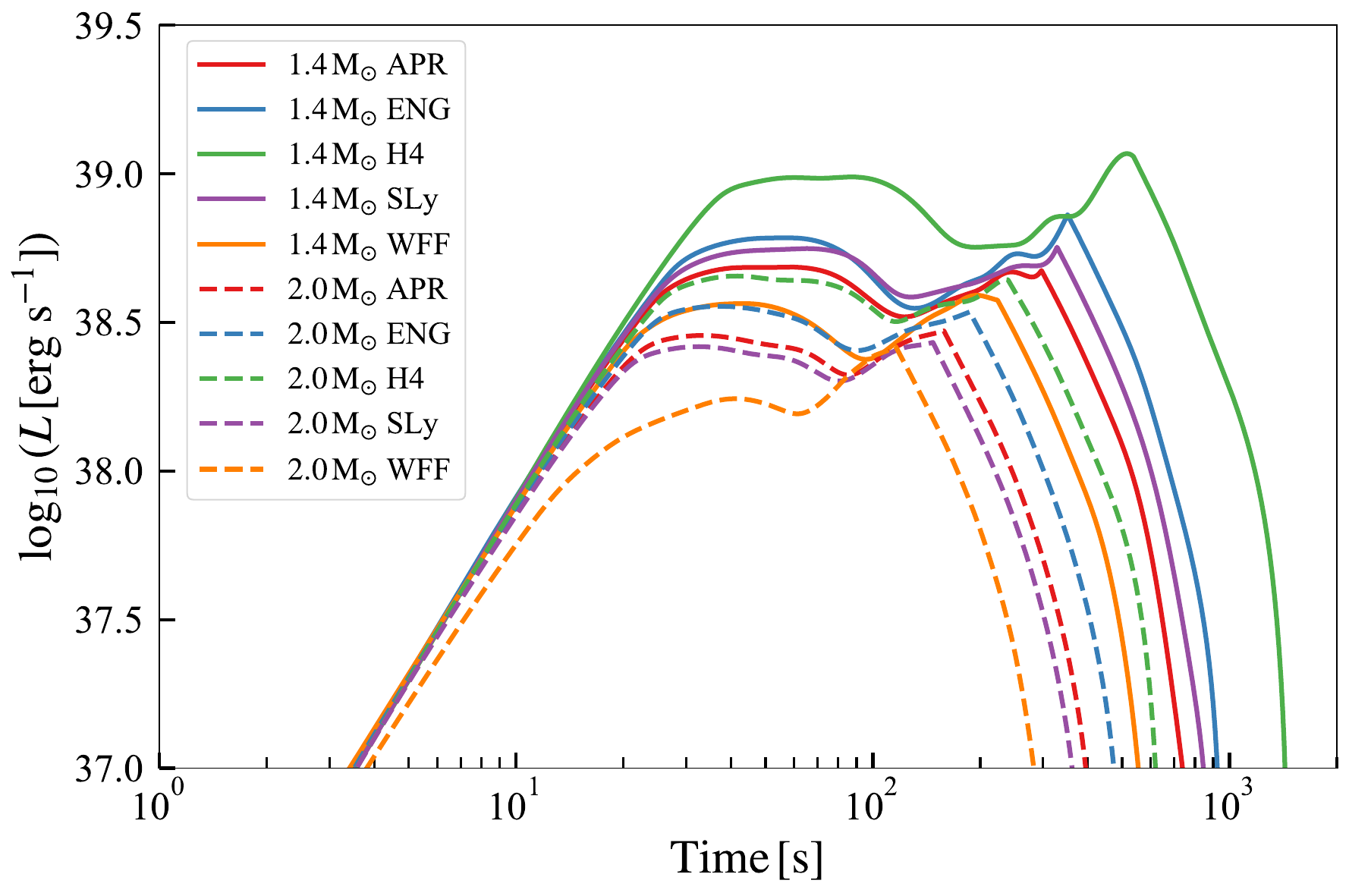}}
    \caption{Time evolution of the bolometric luminosity of novae breves for different EOSs and magnetar masses. Different colors correspond to different EOSs. Solid lines show the results for a $1.4\,\mathrm{M}_\odot$ magnetar, while dashed lines correspond to the $2.0\,\mathrm{M}_\odot$ case. The magnetic field strength is taken to be $B \simeq 10^{15}\,\mathrm{G}$.}
    \label{ fig:lightcurve }
\end{figure}

We present in Fig.~\ref{ fig:lightcurve } the time evolution of the bolometric luminosity for novae breves under different EOSs and magnetar masses. Distinct EOSs lead to noticeable variations in both the peak luminosity and the corresponding peak time. Assuming a fixed magnetar mass of $1.4\,\mathrm{M}_\odot$, we find that the peak bolometric luminosity of novae breves can reach $\simeq 10^{38.5}$--$10^{39}\,\mathrm{erg} \, \mathrm{s}^{-1}$, with peak timescales of $\simeq 10^{2}$--$10^{3}\,\mathrm{s}$. A stiffer EOS (e.g., H4) yields a higher peak luminosity and a longer characteristic timescale. This behavior can be naturally understood in terms of the larger ejecta mass associated with stiffer EOSs (see Table~\ref{ tab:ejecta }).\footnote{In some NS--NS merger scenarios, a stiffer EOS tends to produce less ejecta and power dimmer kilonova emission \citep{Qiumu:2025cyn}, which contrasts with the MGF scenario considered here. This difference arises because stiffer EOSs lead to earlier mergers at larger orbital separations and lower orbital velocities, as well as higher sound speeds that make shock heating of NS matter less efficient.} In addition to the EOS dependence, the magnetar mass also plays an important role in shaping the luminosity evolution of novae breves. As shown in Fig.~\ref{ fig:lightcurve }, a more massive magnetar generally results in a lower peak luminosity and a shorter peak timescale. Figure~\ref{ fig:lightcurve } also indicates a degeneracy between the EOS and the magnetar mass; therefore, additional observational constraints on either parameter would help break this degeneracy and better probe the other. For most light curves, we identify a hump-like feature at $\lesssim 100\,\mathrm{s}$ prior to the luminosity peak. This feature originates from the temporal evolution of the relative recession velocities of the diffusion surface and the photosphere.

We show in Fig.~\ref{ fig:uband } the corresponding $u$-band light-curve evolution, while results for other bands (e.g., $g$, $r$, $i$, $z$) are presented in Fig.~\ref{ fig:muiltiband }. A fiducial Galactic distance of ${D = 10\,\mathrm{kpc}}$ is adopted. The $u$-band light curves exhibit clear EOS-dependent differences in both the minimum apparent AB magnitudes (i.e., at peak brightness), ranging from $\sim 7\,\mathrm{mag}$ for H4 to $\sim 8.5\,\mathrm{mag}$ for WFF, and the peak timescales of $\simeq 10^{2}$--$10^{3}\,\mathrm{s}$. The pronounced peaks allow the peak times to be determined relatively precisely. Similarly to the bolometric case, a more massive magnetar leads to fainter $u$-band emission and a shorter peak timescale. Together, these results suggest that future detections of nova brevis events could provide valuable information on the crustal properties of magnetars.


\subsection{Detection prospects}
\label{ sec:detect }

\begin{figure}[t]
    \centering
    \resizebox{\hsize}{!}{\includegraphics{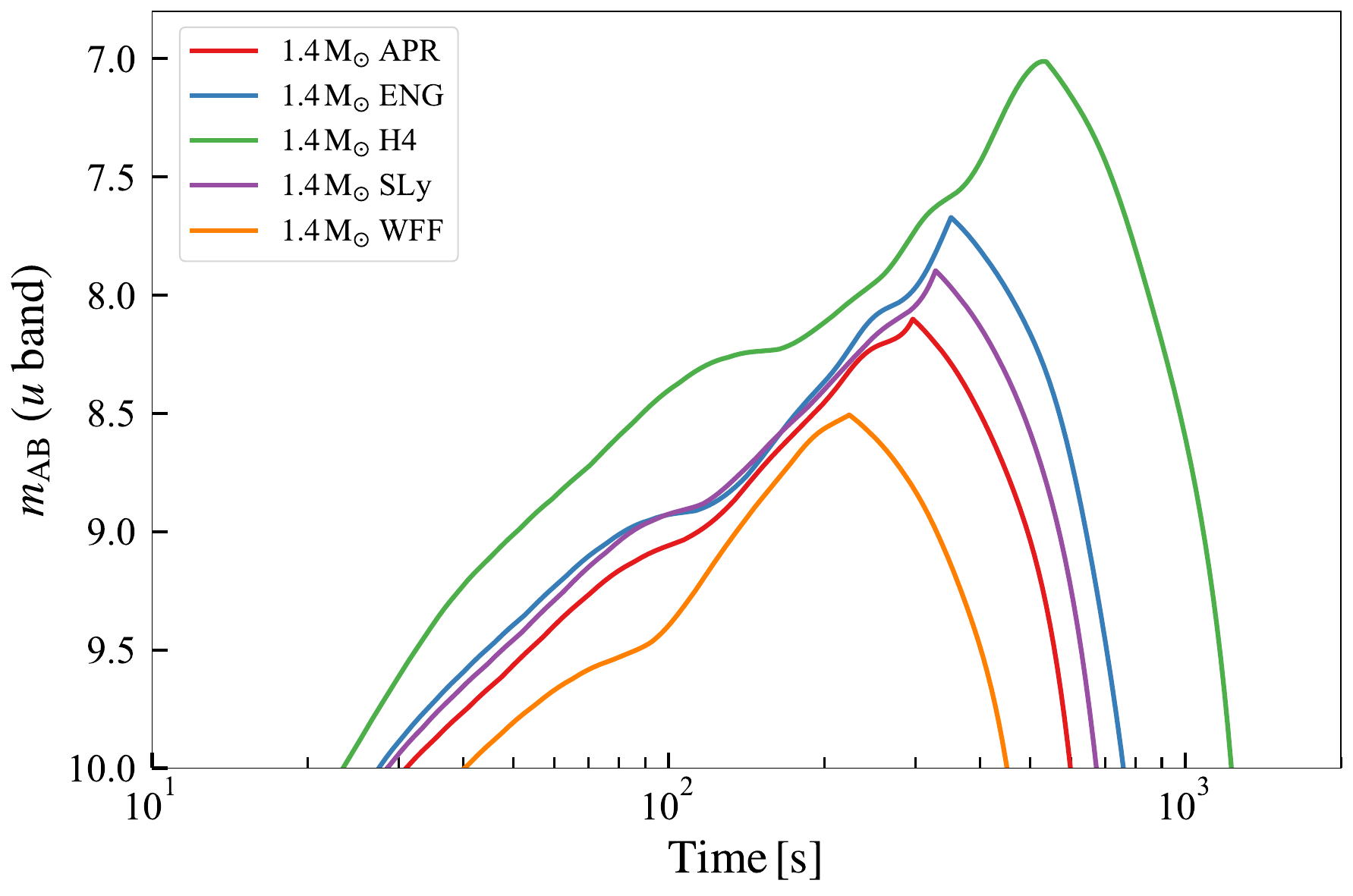}}
    \resizebox{\hsize}{!}{\includegraphics{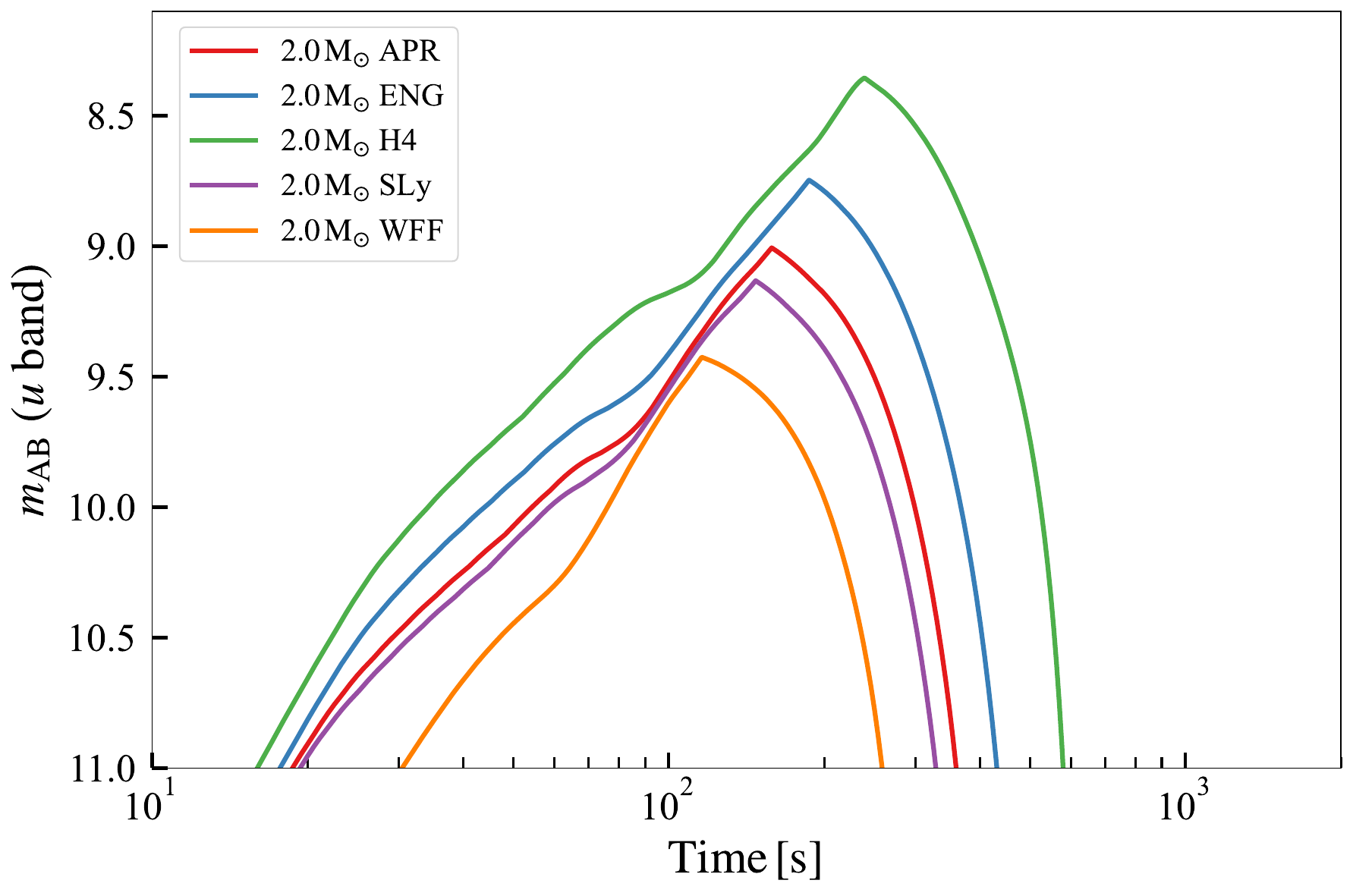}}
    \caption{The $u$-band light-curve evolution of novae breves for different EOSs and magnetar masses. Different colors correspond to different EOSs. The upper panel shows the results for a $1.4\,\mathrm{M}_\odot$ magnetar, while the lower panel corresponds to the $2.0\,\mathrm{M}_\odot$ case. A fiducial Galactic distance of ${D = 10\,\mathrm{kpc}}$ is adopted, and the magnetic field strength is taken to be $B \simeq 10^{15}\,\mathrm{G}$.}
    \label{ fig:uband }
\end{figure}

As shown in Fig.~\ref{ fig:lightcurve } and Fig.~\ref{ fig:uband }, the peak optical luminosity of novae breves is expected to be comparable to that of typical Galactic novae, but with peak timescales of at most several tens of minutes. The detection of such rapidly evolving transients is therefore challenging, although not entirely prohibitive.

For known magnetars, while the occurrence time of potential nova brevis events cannot be predicted in advance, highly sensitive ground-based and space-borne facilities---such as the Large Synoptic Survey Telescope \citep[LSST, now the Vera C. Rubin Observatory;][]{LSSTScience:2009jmu, LSST:2008ijt}---may lead to their discovery. For previously unknown magnetars within or near our Galaxy, instruments with short slewing times or large fields of view can substantially enhance the detection probability. For example, the UltraViolet/Optical Telescope (UVOT) aboard the Neil Gehrels Swift Observatory can repoint to a target within $\lesssim 100\,\mathrm{s}$ following a high-energy trigger \citep{Roming:2005hv}, e.g., from MGFs. Similarly, wide-field survey facilities such as the Wide Field Survey Telescope \citep[WFST;][]{WFST:2023voz} and the China Space Station Telescope \citep[CSST;][]{Wei:2025ycz} also have the potential to detect novae breves associated with MGFs.

To quantitatively assess the detection prospects with current and future facilities, we select magnetars with measured distances and magnetic field strengths from the McGill Magnetar Catalog\footnote{\url{http://www.physics.mcgill.ca/~pulsar/magnetar/main.html}} \citep{Olausen:2013bpa}. In Table~\ref{tab:magnetar_peak_luminosity}, we list the peak bolometric luminosities of the corresponding nova brevis events for different EOSs. Magnetars yielding a total ejecta mass of $M_{\mathrm{ej}} < 10^{-9}\,\mathrm{M}_\odot$ are excluded, as such small ejecta masses are unlikely to support significant nucleosynthesis during a nova brevis event. Assuming a fixed magnetar mass of $1.4\,\mathrm{M}_\odot$, these systems are expected to produce novae breves with peak bolometric luminosities of \mbox{$\sim 10^{37}$--$10^{39}\,\mathrm{erg} \, \mathrm{s}^{-1}$}, comparable to that of typical Galactic novae.\footnote{\citet{Patel:2025tse} reported that SGR 1806$-$20 may eject $10^{-6}\,\mathrm{M}_{\odot}$ of material, with a peak luminosity of $\sim 10^{40}\,\mathrm{erg\,s^{-1}}$ inferred from a delayed MeV signal. Our results are consistent with these findings, since the ejecta mass obtained in our model is smaller, naturally leading to a lower peak luminosity.}

\begin{table*}[t]
	\centering
	\caption{Selected known magnetars and their properties.}
	\renewcommand{\arraystretch}{2}
    \centering
    \setlength{\tabcolsep}{0.5cm}
    {\begin{tabular}{l ccc ccccc}
    \toprule 
    \toprule\\
    [-2.5em]
    Magnetar & $B$ [$\times\,10^{14}$ G] & $D$ [kpc] & \multicolumn{5}{c}{$L_{\rm peak}$ [$\times\,10^{38}$ erg s$^{-1}$]} \\
    \cmidrule(lr){4-8}
     & & & APR & ENG & H4 & SLy & WFF \\
     \midrule
     SGR 1806$-$20 & $19.6$ & 8.7 & $15.3$ & $18.0$ & $28.5$ & $17.2$ & $12.9$ \\
     1E 1841$-$045 & $7.03$ & 8.5 & $3.29$ & $3.87$ & $8.75$ & $3.55$ & $2.69$ \\
     SGR 1900+14 & $7.00$ & 12.5 & $3.27$ & $3.82$ & $8.72$ & $3.55$ & $2.67$ \\
     SGR 0526$-$66 & $5.60$ & 53.6 & $2.69$ & $3.07$ & $5.12$ & $2.95$ & $2.13$ \\
     CXOU J171405.7$-$381031 & $5.01$ & 10.2 & $2.28$ & $2.68$ & $3.64$ & $2.47$ & $1.74$ \\
     1RXS J170849.0$-$400910 & $4.68$ & 3.8 & $2.08$ & $2.42$ & $3.32$ & $2.27$ & $1.57$ \\
     CXOU J010043.1$-$721134 & $3.93$ & 62.4 & $1.47$ & $1.85$ & $2.70$ & $1.72$ & $0.97$ \\
     Swift J1818.0$-$1607 & $3.54$ & 4.8 & $1.14$ & $1.51$ & $2.38$ & $1.33$ & $0.85$ \\
     PSR J1622$-$4950 & $2.74$ & 9.0 & $0.84$ & $0.99$ & $1.48$ & $0.92$ & $0.67$ \\
     SGR J1745$-$2900 & $2.31$ & 8.3 & $0.71$ & $0.83$ & $1.22$ & $0.78$ & $0.49$ \\
     XTE J1810$-$197 & $2.10$ & 3.5 & $0.60$ & $0.75$ & $1.10$ & $0.69$ & $0.33$ \\
     SGR 0501+4516 & $1.87$ & 2.0 & --- & --- & --- & --- & --- \\
     Swift J1834.9$-$0846 & $1.42$ & 4.2 & --- & --- & --- & --- & --- \\
     4U 0142+61 & $1.34$ & 3.6 & --- & --- & --- & --- & --- \\
     Swift J1822.3$-$1606 & $0.14$ & 1.6 & --- & --- & --- & --- & --- \\
     SGR 0418+5729 & $0.06$ & 2.0 & --- & --- & --- & --- & --- \\
     \bottomrule
     \end{tabular}}
     \tablefoot{For each EOS, we assume a fixed magnetar mass of $1.4\,\mathrm{M}_\odot$ when computing the peak bolometric luminosity $L_{\rm peak}$.  Magnetars with a total ejecta mass of $M_{\mathrm{ej}} < 10^{-9}\,\mathrm{M}_\odot$ are not considered, as such small ejecta masses are unlikely to support significant nucleosynthesis during a nova brevis event.}
\label{tab:magnetar_peak_luminosity}
\end{table*}

\begin{figure*}[t]
	\sidecaption
    \includegraphics[width=12cm]{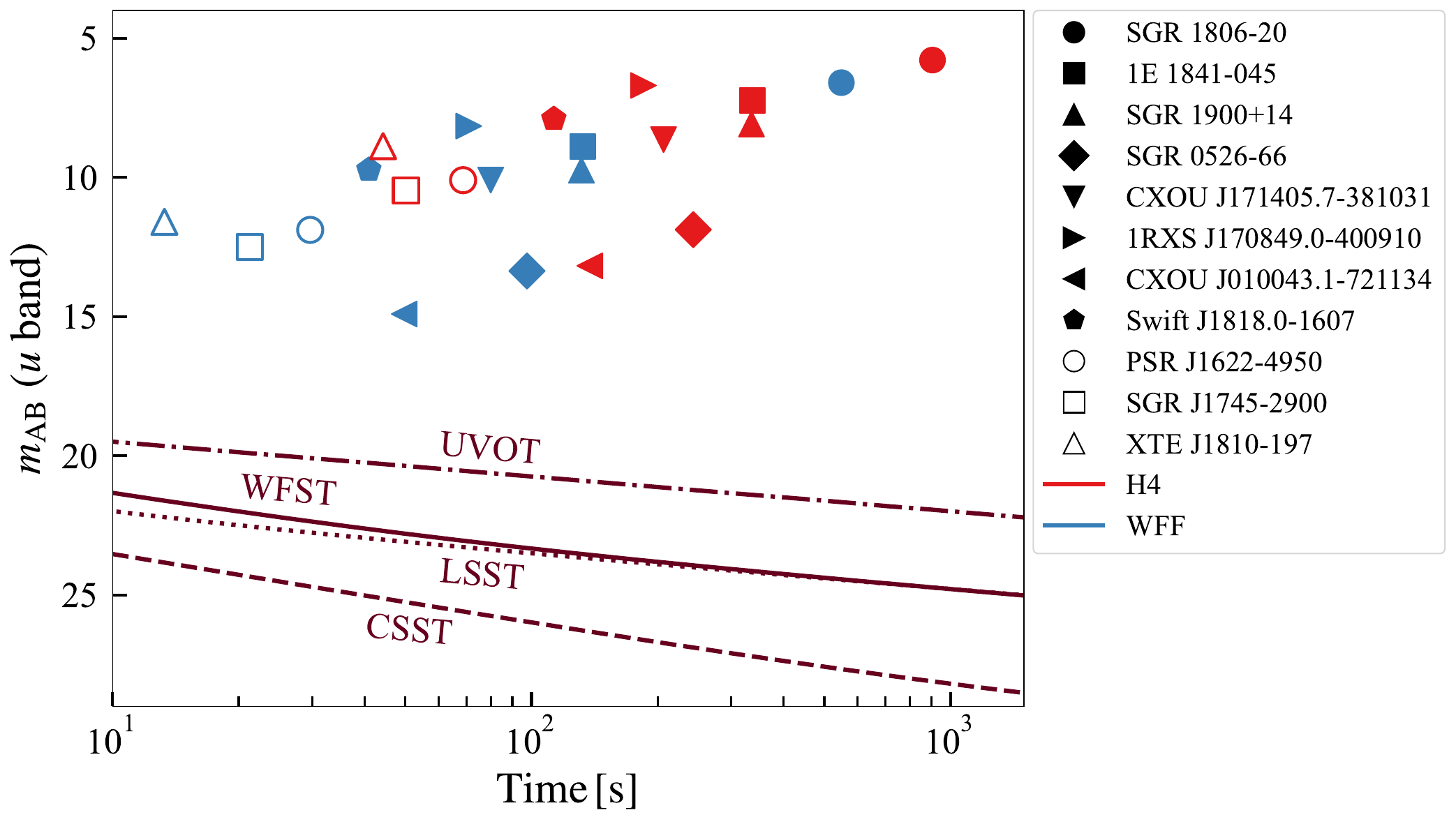}
    \caption{The minimum $u$-band apparent AB magnitudes (i.e., at peak brightness) and the corresponding peak times for selected magnetars in Table~\ref{tab:magnetar_peak_luminosity}. A fixed magnetar mass of $1.4\,\mathrm{M}_\odot$ is adopted. Different symbols represent different magnetars. Among the EOSs considered in this work, the stiffest (H4) and softest (WFF) are indicated by red and blue labels, respectively. The limiting magnitudes as functions of exposure time for different telescopes are overplotted for UVOT, WFST, LSST, and CSST. The region above each line corresponds to the detectable parameter space.}
\label{ fig:u_band_detection }
\end{figure*}

In Fig.~\ref{ fig:u_band_detection }, we show the minimum $u$-band apparent AB magnitudes (at peak brightness) and the corresponding peak times for each selected magnetar. Among the EOSs, the stiffest (H4) and softest (WFF) are indicated by red and blue labels, respectively. We overlay the limiting detection magnitudes as functions of exposure time for LSST\footnote{\url{https://rubin-sim.lsst.io}}, UVOT\footnote{\url{https://www.mssl.ucl.ac.uk/www_astro/uvot/uvot_observing/uvot_tool.html}}, WFST \citep{Lei:2023adp}, and CSST \citep{Wei:2025ycz}, using different line styles. As illustrated, if nova brevis events indeed occur in these magnetar systems, such short-lived transients are potentially detectable with these facilities, particularly when observations are triggered by high-energy alerts from MGFs.

\begin{figure}[t]
	\sidecaption
	\resizebox{\hsize}{!}{\includegraphics{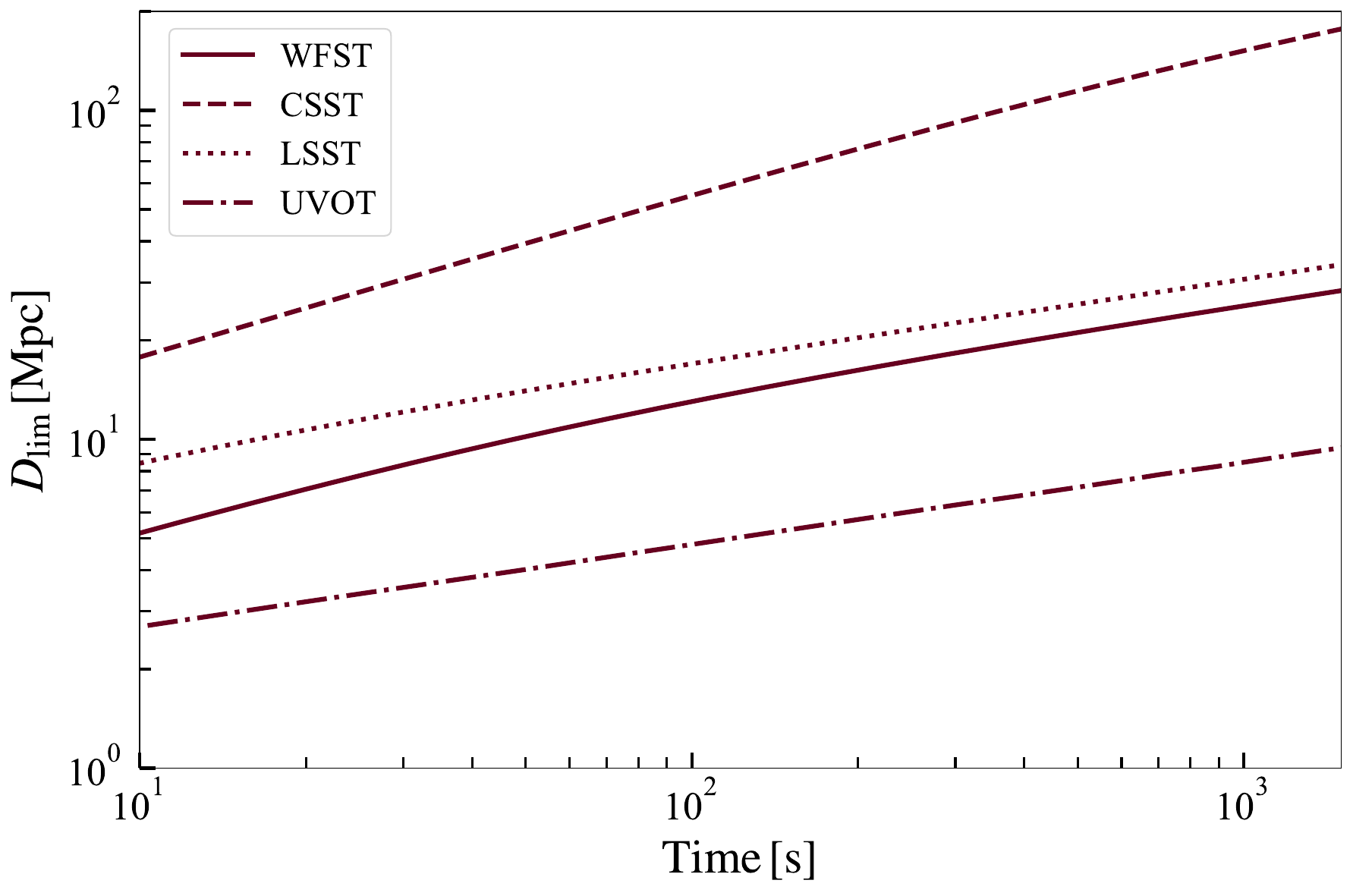}}
    \caption{Detection horizons of novae breves for different facilities, assuming a representative peak optical luminosity of $10^{38}\,\mathrm{erg\,s^{-1}}$. Different linestyles denote different telescopes.}
\label{ fig:detection_horizon }
\end{figure}

Furthermore, assuming a representative peak optical luminosity of $10^{38}\,\mathrm{erg\,s^{-1}}$ for novae breves, we show in Fig.~\ref{ fig:detection_horizon } the corresponding detection horizons for different facilities. A horizon of ${\simeq 10\,\mathrm{Mpc}}$, or even beyond, appears achievable with current and next-generation ground-based and space-borne telescopes. This implies that nearby star-forming galaxies within the Local Volume can be encompassed in systematic searches for nova brevis events.

\section{Conclusion}
\label{ sec:conclu }

Building on previous studies of novae breves powered by the radioactive decay of heavy $r$-process nuclei in MGF ejecta, we have investigated how different NS EOSs and central magnetar masses affect the properties of the ejecta launched following MGFs and the resulting nova brevis emission. Our calculations show that variations in the EOS and magnetar mass can alter the total ejecta mass, as well as its density and velocity distributions, etc., leading to observable differences in the nova brevis light curves. In particular, both the peak luminosity and the characteristic peak timescale are EOS-dependent. Assuming a fixed Galactic magnetar mass of $1.4\,\mathrm{M}_\odot$ and taking the $u$ band as an example, we find that the minimum apparent AB magnitudes (i.e., at peak brightness) range from $\sim 7\,\mathrm{mag}$ for the H4 EOS to $\sim 8.5\,\mathrm{mag}$ for the WFF EOS, with peak timescales of $\simeq 10^{2}$--$10^{3}\,\mathrm{s}$. In addition, a more massive central magnetar generally results in a lower peak luminosity and a shorter peak timescale, indicating a degeneracy between the magnetar mass and the EOS. Detection of such rapidly evolving transients is challenging, but not entirely prohibitive. Their detections may offer an indirect avenue for constraining the magnetar crustal properties. 

We further explore the detection prospects of these fast and faint novae breves with current and next-generation telescopes. Although not every MGF produces an observable nova brevis emission, targeted monitoring provides a promising pathway toward realistic detections from known magnetars, which may generate novae breves with peak bolometric luminosities of $\sim 10^{37}$--$10^{39}\,\mathrm{erg} \, \mathrm{s}^{-1}$, comparable to that of typical Galactic novae. For unknown magnetars, instruments with rapid slewing capabilities and wide fields of view offer a complementary advantage in capturing such short-lived nova brevis transients. A detection horizon of ${\simeq 10\,\mathrm{Mpc}}$, or even beyond, achievable with current and future facilities, would encompass the nearest star-forming galaxies within the Local Volume. 

It is important to note that we adopt the mass ejection model proposed by \citet{Cehula:2023hdh}, and our results therefore apply specifically within the framework of that study. The exact mass ejection mechanism, however, remains uncertain. For example, \citet{Bransgrove:2025tpa} performed two-dimensional magnetohydrodynamic simulations of a markedly different scenario in which baryon ejection is driven by the eruption of baryon-loaded magnetic loops from the NS interior. These differences highlight the need for further investigation of the mass ejection process. Overall, our results indicate that future searches for novae breves are feasible. Continued theoretical and observational efforts will be essential to confirm such transients and deepen our understanding of NS physics and the origin of heavy elements.


\begin{acknowledgements}
\label{ sec:acknow }

We thank the anonymous referee for helpful comments. This work is supported by the Beijing Natural Science Foundation (QY25099, 1242018),  the National Natural Science Foundation of China (12573042, 12403043, and 12473038), the National SKA Program of China (2020SKA0120300), the Max Planck Partner Group Program funded by the Max Planck Society, and the High-Performance Computing Platform of Peking University.

\end{acknowledgements}


\bibliographystyle{aa}
\bibliography{refs} 

\clearpage


\begin{appendix}

\section{Overview of the unbound ejecta model}
\label{ app:A }

For a given EOS and magnetar mass $M$, we compute the stellar radius $R$ as well as the internal pressure and density profiles, $P(r)$ and $\rho(r)$, respectively \citep[see e.g.,][]{Lattimer:2000nx, Ozel:2016oaf, Lattimer:2021emm, Gao:2021uus}. Within the magnetar crust, the condition of hydrostatic equilibrium gives
\begin{equation}
    \dfrac{1}{\rho_{\mathrm{cr}}}\dfrac{\mathrm{d}P_{\mathrm{cr}}}{\mathrm{d}r} = -g(r) \simeq -G\dfrac{M}{R^2}\,,
\label{ eq:hydrostatic equilibrium }
\end{equation}
where $G$ is the gravitational constant, and $g(r)$ is the local gravitational acceleration, which is approximated as constant across the relatively thin crustal layer. Under this approximation, the mass of crustal material enclosed above a given radius $r$ is
\begin{equation}
    M_{\mathrm{cr}}(>r) = \int_r^{R} 4\pi r'^2 \rho_{\mathrm{cr}}(r') \mathrm{d}r' \simeq \dfrac{4\pi R^4}{GM} P_{\mathrm{cr}}(r)\,.
\label{ eq:M_P }
\end{equation}
Therefore, specifying the inner boundary of the unbound region---equivalently, the critical pressure or density at that boundary---uniquely determines the total mass of the unbound ejecta.

Following \citet{Patel:2025frn, Patel:2025tse}, we consider that the energy released during a MGF not only generates a pair-photon fireball above the magnetar surface, but also drives a strong radiation-dominated shock into the magnetar crust. For a typical magnetar, the magnetic field strength in the magnetosphere is \mbox{$B \simeq 10^{14}$--$10^{15}\,\mathrm{G}$}. The external pressure exerted on the crust during MGFs can therefore be approximated as \mbox{$P_{\mathrm{ext}} \gtrsim B^2 / 8\pi$}. For a shocked crustal layer to ultimately escape from the star, \citet{Patel:2025frn, Patel:2025tse} derived the maximum crustal density, corresponding to the innermost unbound layer, as
\begin{equation}
	\rho_{\mathrm{cr,max}} \simeq \dfrac{4P_{\mathrm{ext}}R}{7GM}\,.
\label{ eq:unbound }
\end{equation}
Combining Eq.~\eqref{ eq:M_P } and Eq.~\eqref{ eq:unbound }, together with the pressure--density relation $P(\rho)$, the total unbound ejecta mass $M_{\mathrm{ej}}$ can be calculated for different EOSs and magnetar masses. The resulting values are listed in Table~\ref{ tab:ejecta }.

Assuming homologous expansion, $R_{\mathrm{ej}} \propto \varv t$, the volume element of each ejecta layer scales as $\mathrm{d}V \propto R_{\mathrm{ej}}^2\mathrm{d}R_{\mathrm{ej}} \propto \varv^2\mathrm{d}\varv$. Motivated by hydrodynamical simulations \citep{Cehula:2023hdh}, the unbound ejecta are assumed to follow a distribution of asymptotic velocities given by 
\begin{equation}
    \frac{d m}{d \varv} \propto \varv^{-\alpha}\,,
\label{ eq:mass distribution }
\end{equation}
where we adopt the Lagrangian mass coordinate ${m \equiv M_{\mathrm{cr}}(>r)}$. In this work we take $\alpha \simeq 4$ \citep{Patel:2025frn, Patel:2025tse}. After normalization, the velocity profile of the ejecta can be written as
\begin{equation}
    \varv = \varv_{\mathrm{i}} \left( \dfrac{m}{M_{\mathrm{ej}}} \right)^{-1/3}\,,
\label{ eq:v_m }
\end{equation}
where ${\varv_{\mathrm{i}}}$ denotes the velocity of the innermost ejecta layer. We adopt $\varv_{\mathrm{i}} \simeq 0.2c$ with the speed of light $c$, consistent with  \citet{Cehula:2023hdh} and \citet{Patel:2025frn, Patel:2025tse}. The corresponding density profile can then be estimated as ${\rho \simeq \mathrm{d}m / \mathrm{d}V \propto \varv^{-6}}$. After normalization, this yields
\begin{equation}
    \rho = \dfrac{3}{4\pi} \dfrac{M_{\mathrm{ej}}}{({\varv_{\mathrm{i}}}t)^3} \left( \dfrac{{\varv_{\mathrm{i}}}}{\varv} \right)^6\,.
\label{ eq:rho_homo }
\end{equation}
As discussed by \citet{Patel:2025frn, Patel:2025tse}, Eq.~\eqref{ eq:rho_homo } applies during the late-time three-dimensional expansion phase. At earlier times, the ejecta evolution is better approximated as a planar one-dimensional expansion with constant velocity. Accordingly, the Lagrangian density evolution can be written as
\begin{equation}
    \rho(t) = \rho_{\mathrm{D}} 
    \begin{cases}
    \left(\dfrac{t}{t_{\mathrm{D}}}\right)^{-1}, & t_{0} \leq t \leq t_{\mathrm{D}} \\
    \left(\dfrac{t}{t_{\mathrm{D}} }\right)^{-3}, & t \geq t_{\mathrm{D}} \,,
    \end{cases}
\label{ eq:rho_evolve }
\end{equation}
where $t_0$ denotes the start time of the nucleosynthesis (see Appendix~\ref{ app:B }), and $t_{\mathrm{D}} \equiv 2R/\varv$ is the time at which each ejecta layer has expanded by a distance comparable to the magnetar diameter. Using Eq.~\eqref{ eq:rho_homo }, the density at $t_{\mathrm{D}}$ is ${\rho_{\mathrm{D}} = 3 M_{\mathrm{ej}} \varv_{\mathrm{i}}^3 / (4\pi t_{\mathrm{D}}^3 \varv^6)}$. The initial density is then $\rho_0 =\rho_{\mathrm{D}} t_{\mathrm{D}} / t_0$.

Moreover, the temperature of each ejecta layer at the start time, $T_{0}=T\left(t_{0}\right)$, is determined by the initial density $\rho_0$ and the specific entropy. Following \citet{Patel:2025frn, Patel:2025tse}, we adopt the immediate post-shock density $\rho_{\mathrm{sh}} \approx 7 \rho_{\mathrm{cr}}$ and the specific internal energy $e_{\mathrm{sh}} \approx 3 P_{\mathrm{ext}} / \rho_{\mathrm{sh}}$, as implied by the Rankine--Hugoniot jump conditions \citep[see e.g., Eq.~(3) and Eq.~(4) in][]{Patel:2025tse}. These quantities are used to infer the specific entropy, and hence the initial temperature distribution of the unbound ejecta, which serves as input for the subsequent nucleosynthesis calculations performed with \texttt{SkyNet} \citep{Lippuner:2015gwa, Lippuner:2017tyn}. Readers interested in more technical details are referred to Appendix~A of \citet{Patel:2025tse}.


\section{Nucleosynthesis and light-curve calculations}
\label{ app:B }

\begin{figure*}[ht]
    \centering
    \includegraphics[width=17cm]{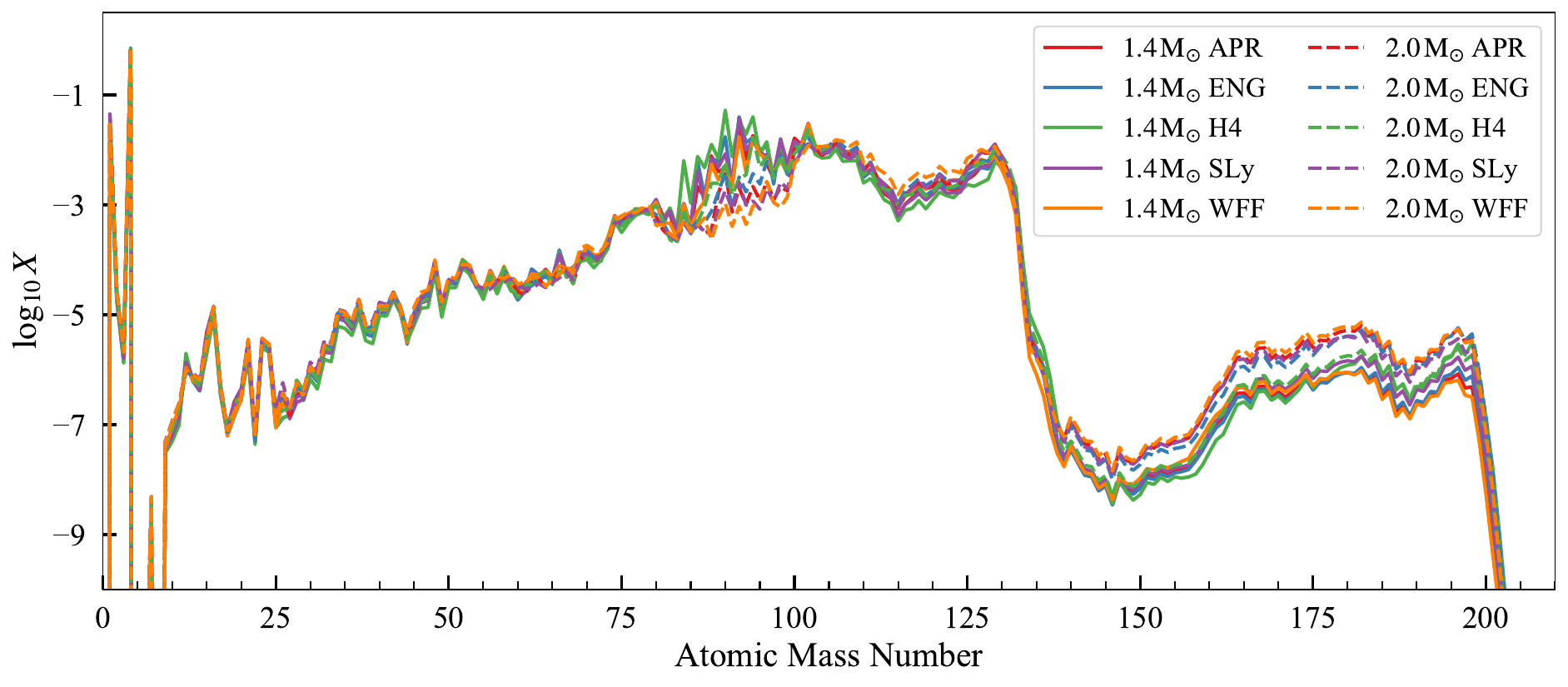}
    \caption{Total nucleosynthesis yields of the ejecta computed with \texttt{SkyNet} for different EOSs and magnetar masses. The magnetic field strength is taken to be $B \simeq 10^{15}\,\mathrm{G}$. }
\label{ fig:abundance }
\end{figure*}

\begin{figure*}[ht]
    \centering
    \includegraphics[width=8.25cm]{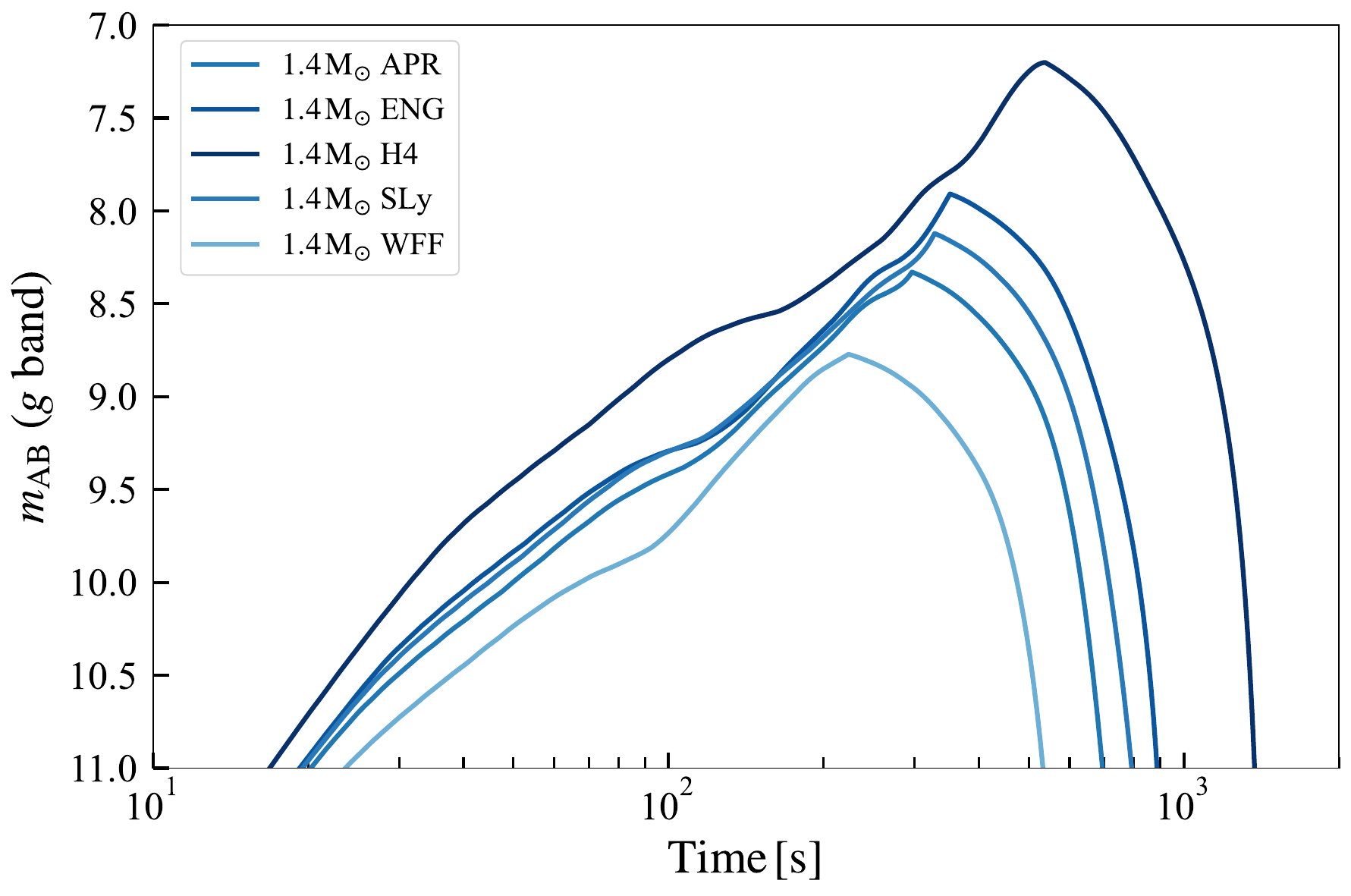}
    \hspace{0.5cm}
    \includegraphics[width=8.25cm]{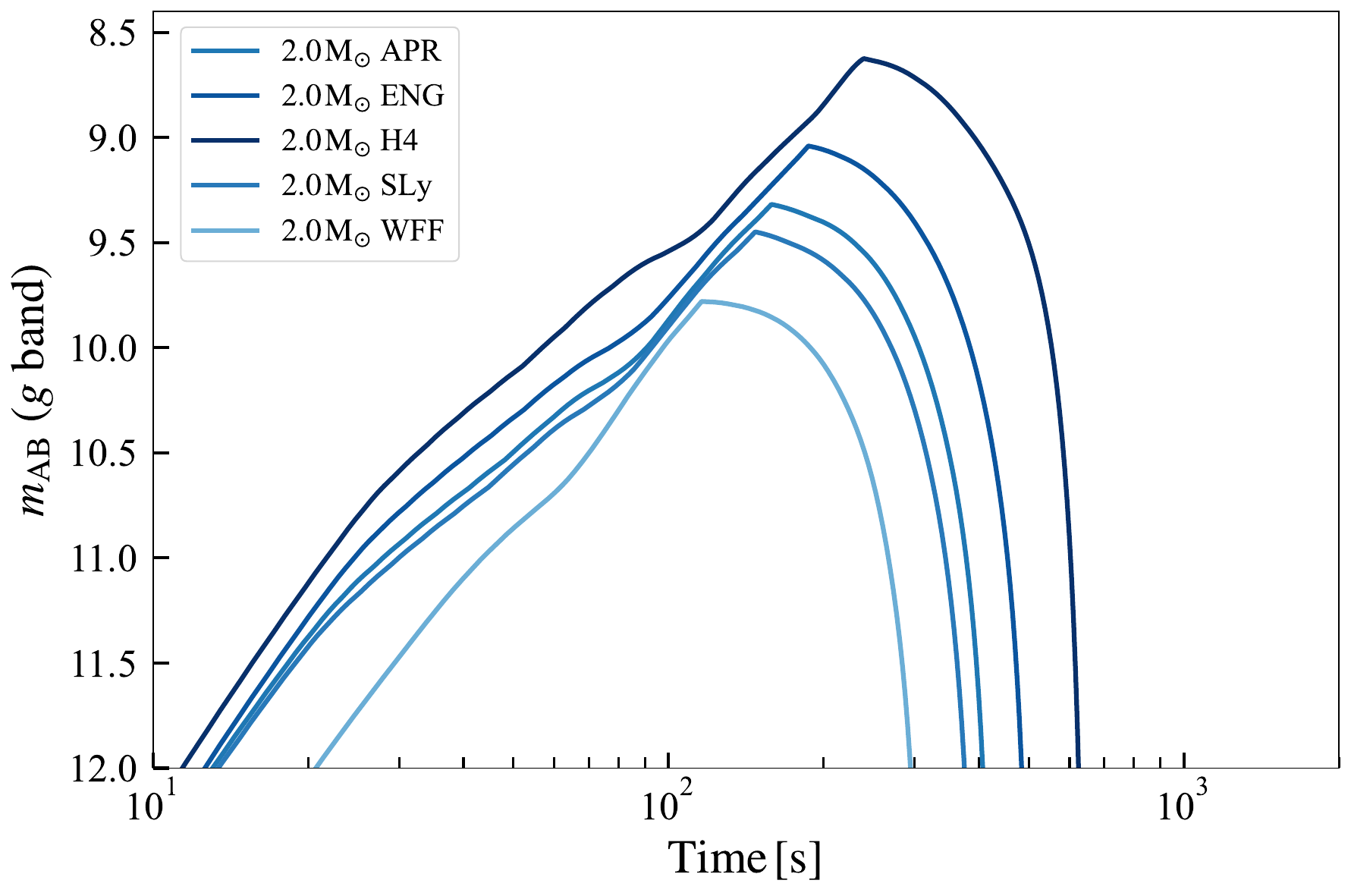}
    
    \includegraphics[width=8.25cm]{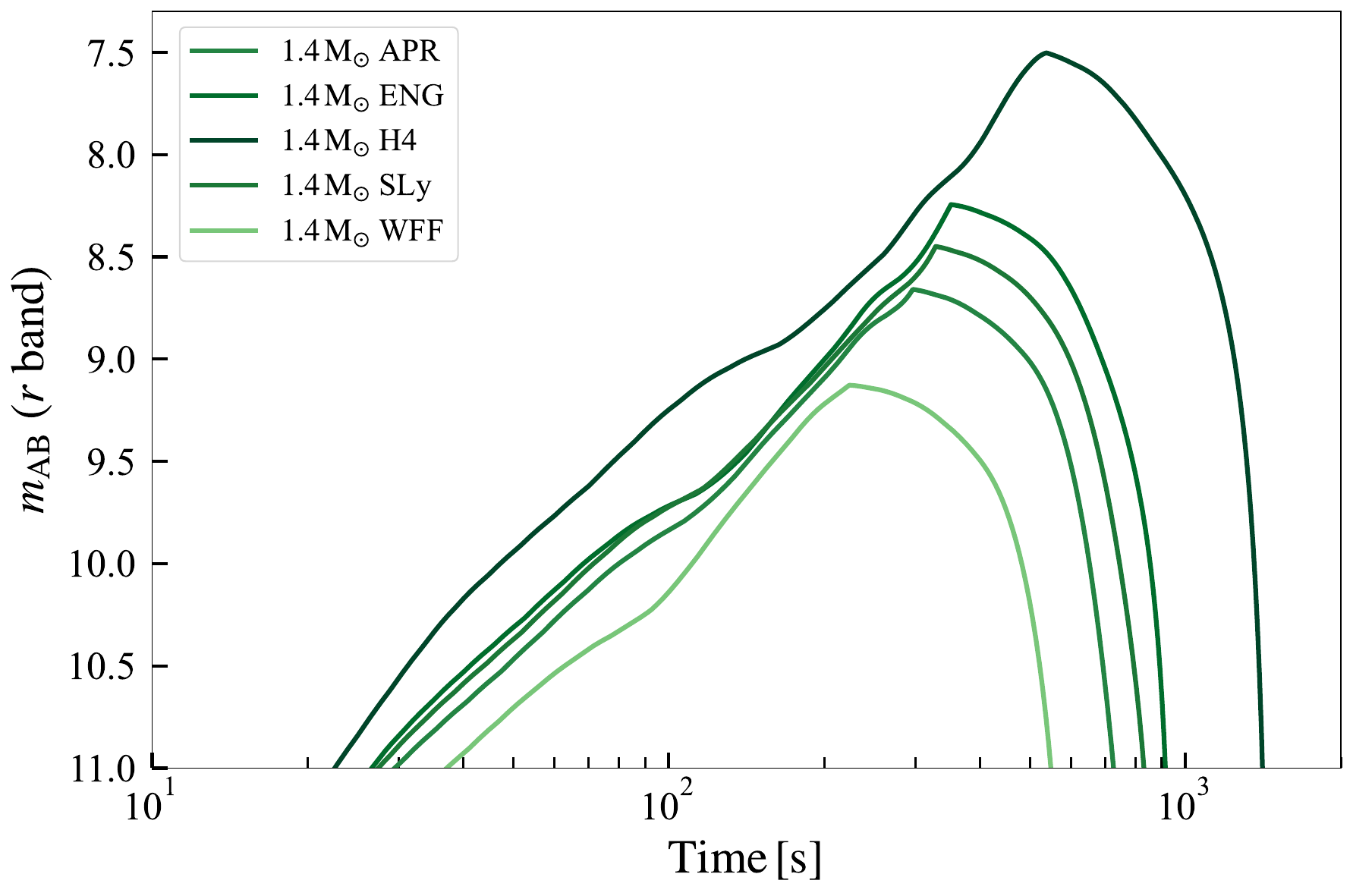}
    \hspace{0.5cm}
    \includegraphics[width=8.25cm]{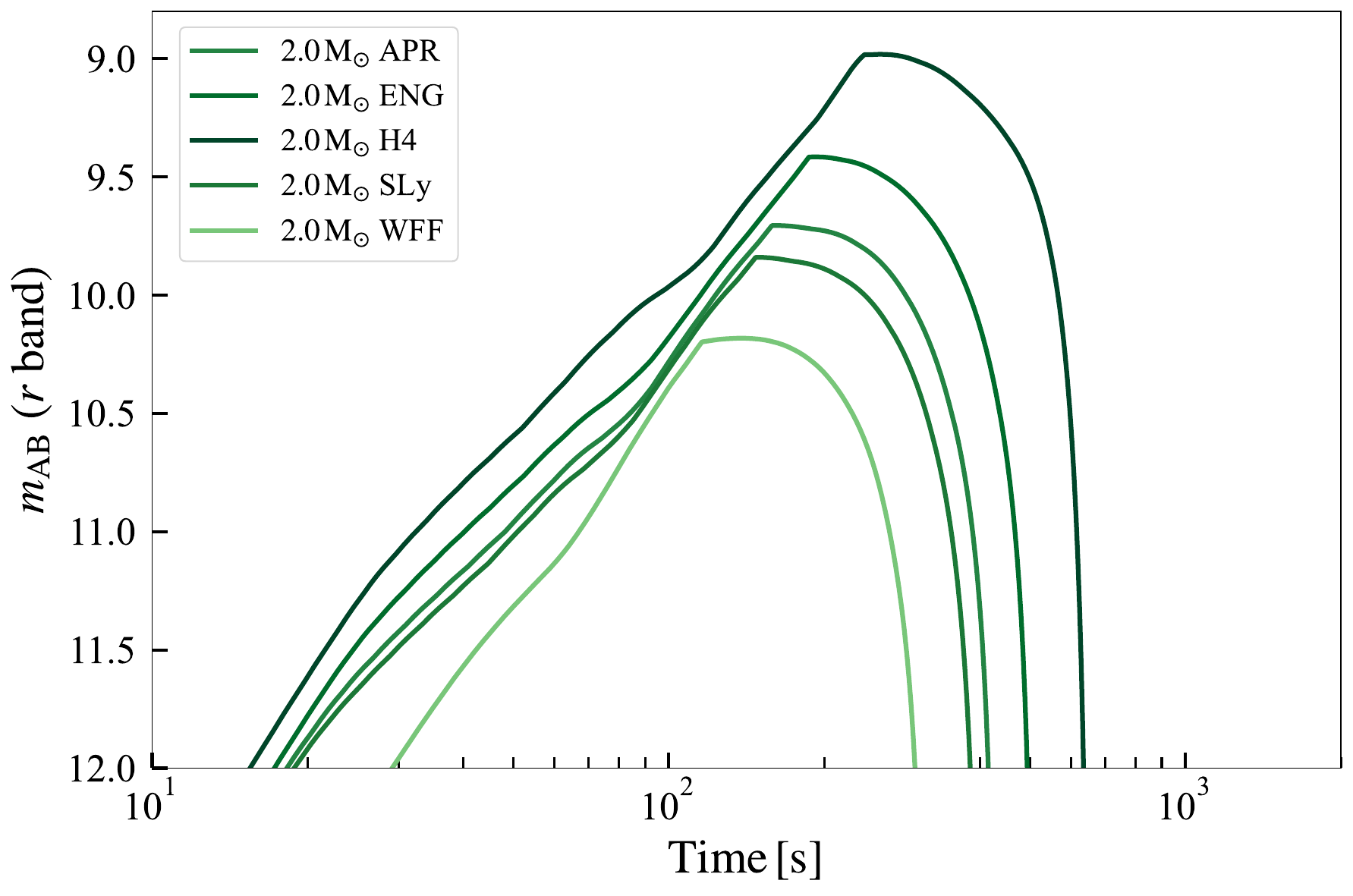}
    
    \includegraphics[width=8.25cm]{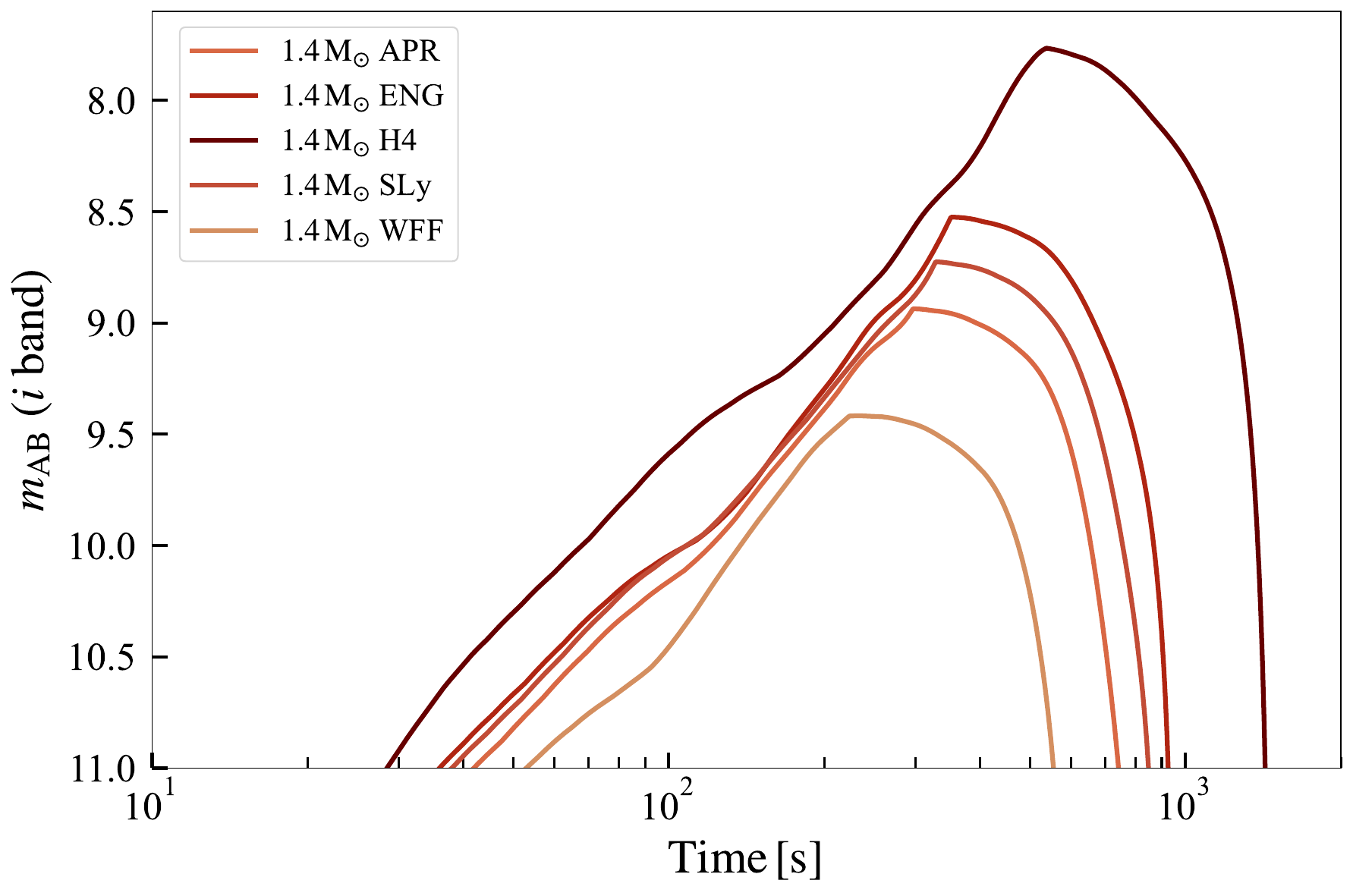}
    \hspace{0.5cm}
    \includegraphics[width=8.25cm]{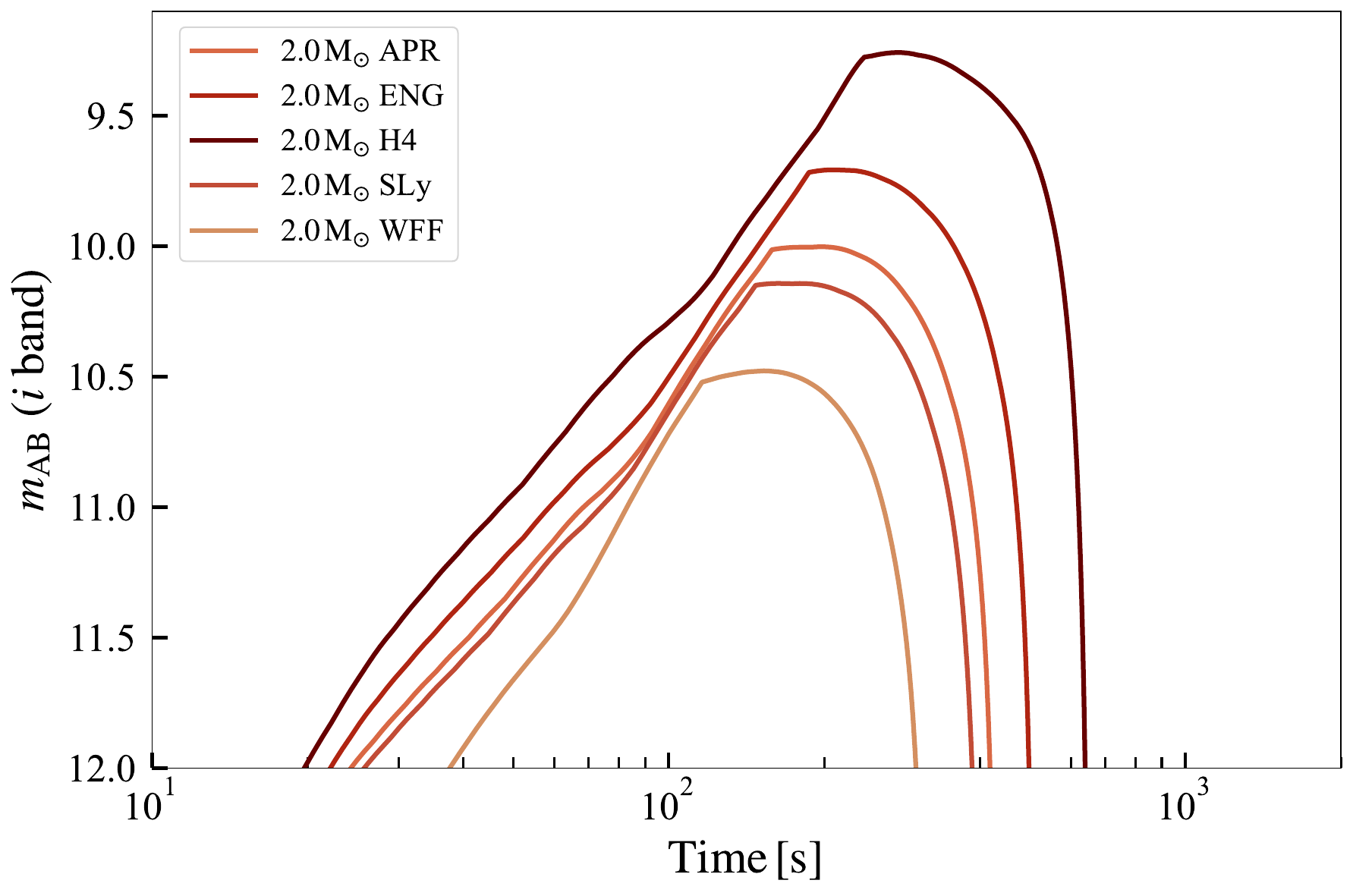}
    
    \includegraphics[width=8.25cm]{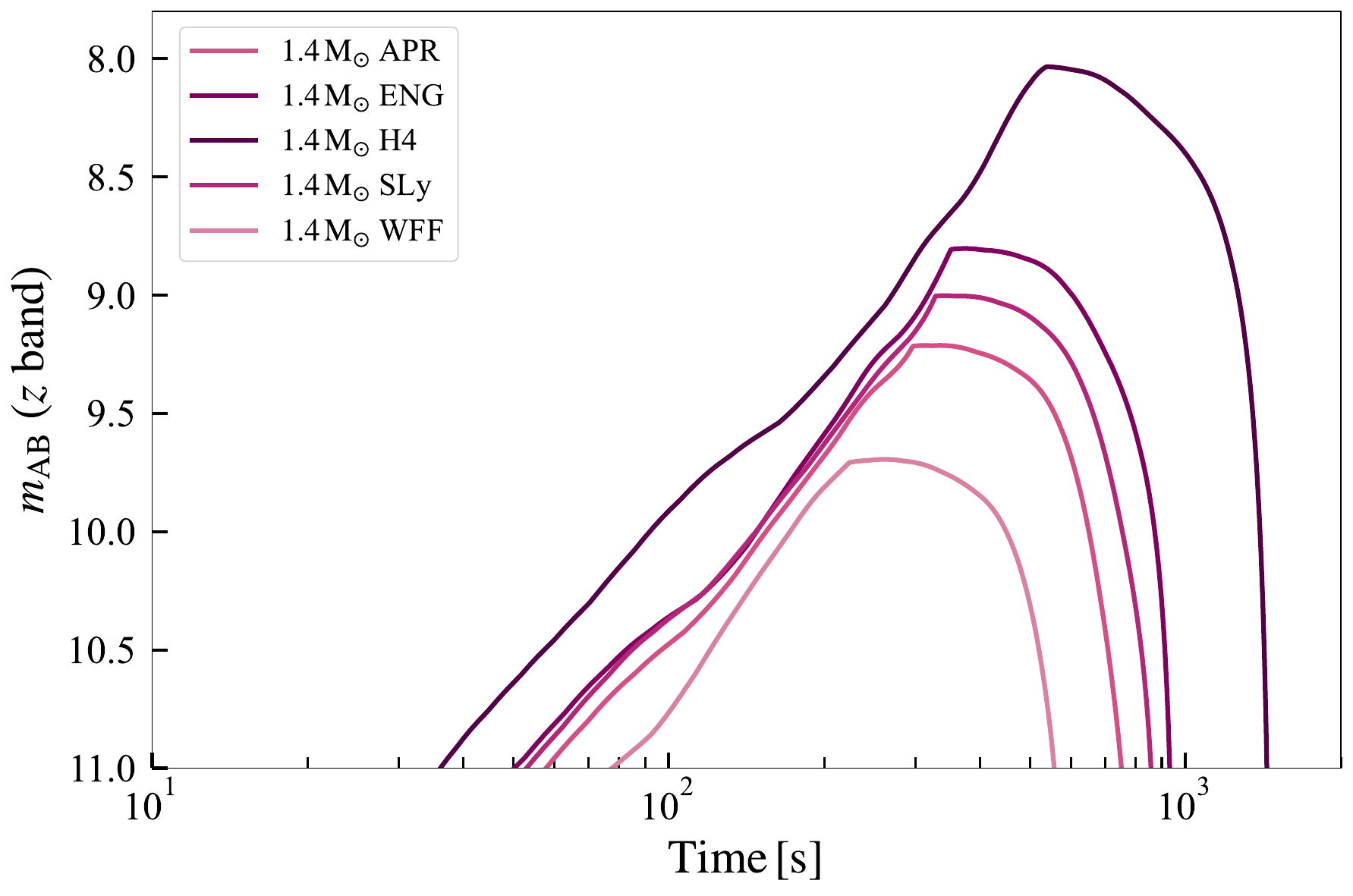}
    \hspace{0.5cm}
    \includegraphics[width=8.25cm]{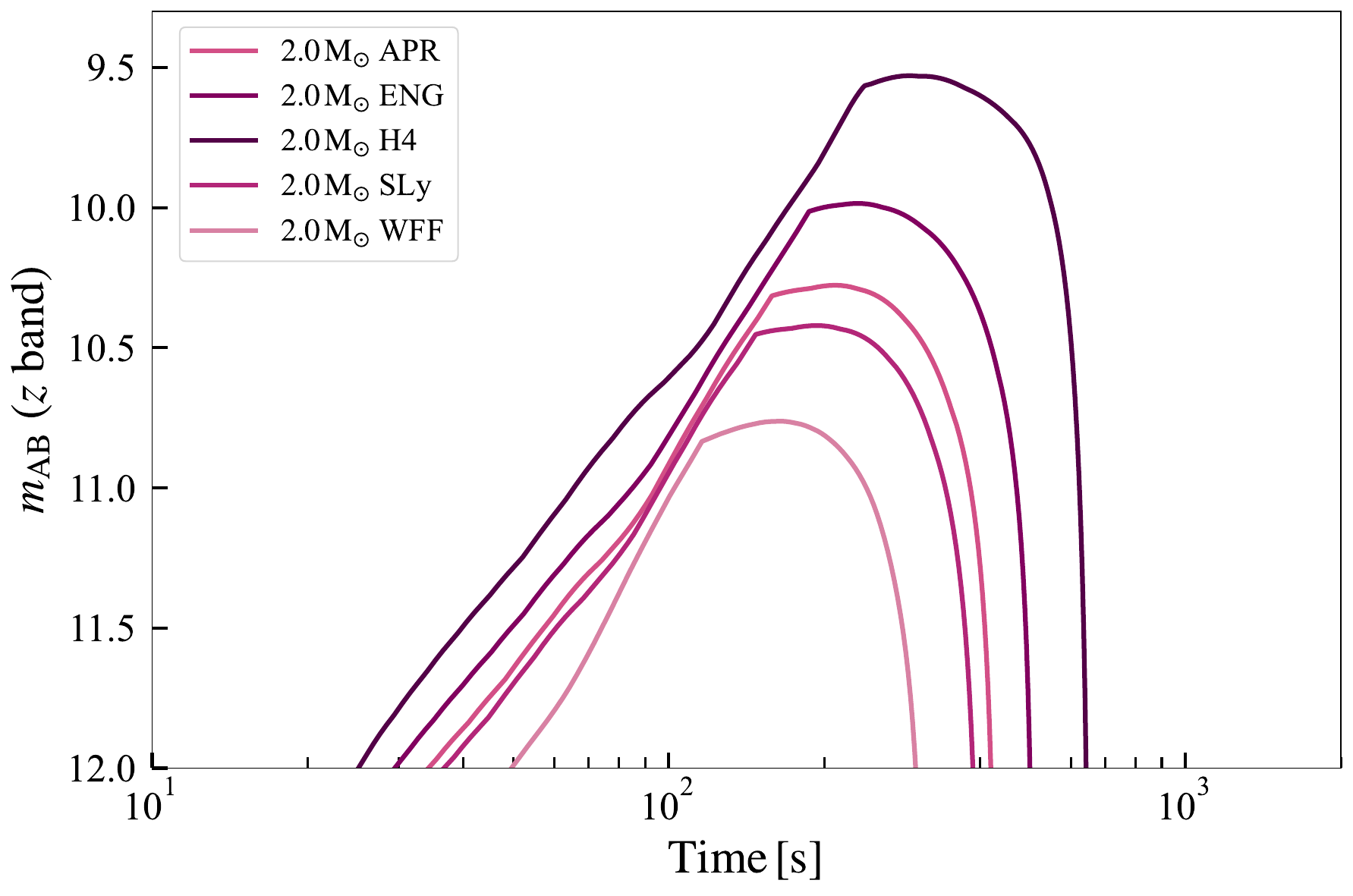}
    \caption{Similar to Fig.~\ref{ fig:uband }, but for (from top down) $g$, $r$, $i$, and $z$ bands. Different colors correspond to different EOSs. The left panels show the results for a $1.4\,\mathrm{M}_\odot$ magnetar, while the right panels correspond to the $2.0\,\mathrm{M}_\odot$ case. A fiducial Galactic distance of ${D = 10\,\mathrm{kpc}}$ is used.}
\label{ fig:muiltiband }
\end{figure*}

For each adopted NS EOS and magnetar mass, we perform nucleosynthesis calculations using the \texttt{SkyNet} reaction network \citep{Lippuner:2015gwa, Lippuner:2017tyn}, from which we obtain the time evolution of nuclear abundances and the corresponding total elemental yields of the ejecta. The resulting total abundance patterns are compared in Fig.~\ref{ fig:abundance }

Following \citet{Patel:2025frn, Patel:2025tse}, the bolometric luminosity of novae breves is primarily determined by the radioactive energy deposited within the optically thick ejecta. Using the Lagrangian mass coordinate $m \leqslant M_{\mathrm{ej}}$, where $m = 0$ corresponds to the outermost layer of the ejecta, the time-dependent luminosity can be expressed as
\begin{equation}
    L_{\mathrm{ph}}(t) = \int_{M_{\mathrm{ph}}}^{M_{\mathrm{diff}}} \dot{q}(t) \, \mathrm{d}m \,,
\label{ eq:bolometric luminosity }
\end{equation}
where $M_{\mathrm{ph}}$ and $M_{\mathrm{diff}}$ are the mass coordinates of the photosphere and the diffusion surface, respectively, and $\dot{q}(t)$ denotes the specific heating rate. The photosphere is defined as the layer at which the optical depth satisfies $\tau \simeq 2/3$, while the diffusion surface corresponds to $\tau = c/\varv$. Above the diffusion surface, the photon diffusion timescale becomes shorter than the dynamical expansion timescale, allowing photons to escape efficiently without significant adiabatic losses. As a result, the region bounded by the diffusion surface and the photosphere provides the dominant contribution to the bolometric luminosity. Notably, the optical depth is computed as $\tau = \int^{\infty}_{R_{\mathrm{ej}}} \kappa  \rho \, \mathrm{d}R_{\mathrm{ej}}$, where the opacity evolves with time according to \citep{Patel:2025tse},
\begin{equation}
	\begin{aligned}
    	\kappa(t) = & \ X_{\mathrm{p}}(t) \kappa_{\mathrm{p}} + X_{\alpha}(t) \kappa_{\alpha} \\
    	            &  + X_{\mathrm{seed}}(t) \kappa_{\mathrm{seed}} + X_{\mathrm{2nd}}(t) \kappa_{\mathrm{2nd}} + X_{\mathrm{3rd}}(t) \kappa_{\mathrm{3rd}}\,,
    \end{aligned}
\label{ eq:opacity }
\end{equation}
where $X_i(t)$ represents the mass fraction of the $i$-th species. The adopted component opacities are $\kappa_{\mathrm{p}} \approx 0.38\, \mathrm{cm^2\,g^{-1}}$ for protons, $\kappa_{\alpha} \approx 0.2\, \mathrm{cm^2\, g^{-1}}$ for alpha particles, $\kappa_{\mathrm{seed}} \approx 0.5\, \mathrm{cm^2\, g^{-1}}$ for light seed nuclei, $\kappa_{\mathrm{2nd}} \approx 3\, \mathrm{cm^2\, g^{-1}}$ for second-peak $r$-process nuclei, and $\kappa_{\mathrm{3rd}} \approx 20\, \mathrm{cm^2\, g^{-1}}$ for third-peak $r$-process nuclei. 

The total specific heating rate $\dot{q}(t)$ is computed by summing the contributions from $\alpha$, $\beta$, and $\gamma$ decays occurring in the unbound ejecta, i.e.,
\begin{equation}
	\dot{q}(t) = \sum_{\mathrm{n}}\,f_{\mathrm{n}}(t)\, \dot{q}_{\mathrm{n}}(t)\,, \quad \mathrm{n} \in \{ \alpha ,\, \beta,\,\gamma\} \,,
\label{ eq:specific heating rate }
\end{equation}
where $f_{\mathrm{n}}(t)$ denotes the thermalization efficiency of each decay channel. The specific radioactive heating rate $\dot{q}_{\mathrm{n}}(t)$ is obtained by summing over all unstable nuclides in the ejecta,
\begin{equation}
    \dot{q}_{\mathrm{n}}(t) = N_{\mathrm{A}} \sum_\mathrm{i} \lambda_i Y_i(t) E_{\mathrm{n}}\,, \quad \mathrm{n} \in \{\alpha, \beta, \gamma\}\,,
\label{ eq: specific radioactive decay energy }
\end{equation}
where $N_{\mathrm{A}}$ is Avogadro's number, $\lambda_i = \ln 2 / T_{1/2\,,i}$ is the decay rate of the $i$-th nuclide with half-life $T_{1/2\,,i}$, $Y_i(t)$ is its abundance, and $E_{\mathrm{n}}$ is the energy released per decay in channel $\mathrm{n}$. For heavy $r$-process nuclei, the decay energies are taken from the Evaluated Nuclear Data File library \citep[ENDF/B-VIII.1;][]{Nobre:2025xlg}. 

For the thermalization efficiency $f_{\mathrm{n}}(t)$, we adopt the analytic prescriptions of \citet{Kasen:2018drm}. For $\beta$ decay, we use
\begin{equation}
    f_{\beta}(t) = \left(1 + \dfrac{t}{t_{\beta}}\right)^{-1}\,,
\label{ eq:fbeta }
\end{equation}
where the characteristic thermalization timescale $t_{\beta}$ is given by
\begin{equation}
    t_{\beta} \simeq 2 \left(\frac{M_{\mathrm{ej}}}{10^{-8}\,\mathrm{M}_\odot}\right)^{2/3} \left(\frac{\varv_{\mathrm{ej}}}{0.2\,c}\right)^{-2} \zeta^{2/3} \,\mathrm{min}\,,
\label{ eq:tbeta }
\end{equation}
with $M_{\mathrm{ej}}$ the total ejecta mass and $\varv_{\mathrm{ej}}$ the characteristic ejecta velocity. $\zeta$ is a dimensionless constant, for which we adopt $\zeta \simeq 1$ \citep{Chen:2023cco}. For $\alpha$ decay, the thermalization efficiency is approximated as
\begin{equation}
    f_{\alpha}(t) = \left(1 + \dfrac{t}{t_{\alpha}}\right)^{-1.5}\,,
\label{ eq:falpha }
\end{equation}
where the corresponding thermalization timescale is $t_{\alpha} \simeq 3t_{\beta}$. For $\gamma$ emission, we adopt
\begin{equation}
    f_{\gamma}(t) = 1 - e^{- t_{\gamma}^2 / t^2}\,,
\label{ eq:fgamma }
\end{equation}
with the thermalization timescale,
\begin{equation}
    t_{\gamma} = \left( \frac{3\kappa_\gamma M_{\mathrm{ej}}}{16\pi \varv_{\mathrm{ej}}^2} \right)^{1/2}\,,
\label{ eq:tgamma }
\end{equation}
where we adopt a $\gamma$-ray opacity of $\kappa_\gamma \simeq 0.02~\mathrm{cm^2\,g^{-1}}$.

The monochromatic flux density at an observing frequency $\nu$ is given by
\begin{equation}
    F_{\nu}=\frac{2\pi h{\nu}^3}{c^2}\,\left[{\mathrm{exp}(\dfrac{\,h\nu}{kT_{\mathrm{ph}}})-1}\right]^{-1}\,\frac{R_{\mathrm{ph}}^2}{D^2}\,, 
\end{equation}
where $h$ is the Planck constant, $R_{\mathrm{ph}}$ is the photospheric radius, $D$ is the luminosity distance, and $T_{\mathrm{ph}}$ denotes the photospheric temperature. Using Eq.~(\ref{ eq:bolometric luminosity }), the photospheric temperature is estimated as
\begin{equation}
    T_{\mathrm{ph}}=\left(\dfrac{L_{\mathrm{ph}}}{4\pi\sigma R^2_{\mathrm{ph}}}\right)^{1/4}\,,
\end{equation}
where $\sigma$ is the Steffan--Boltzmann constant. The corresponding AB magnitude is then computed as ${m_{\mathrm{AB}}= -2.5\,\mathrm{log}_{10}\left({F_{\nu}}\,/\,{3631\,\mathrm{Jy}}\right)}$. 

In addition to the $u$-band light-curve evolution of novae breves for different EOSs and magnetar masses shown in Fig.~\ref{ fig:uband }, we also present their corresponding light curves in other optical bands ($g$, $r$, $i$, $z$) in Fig.~\ref{ fig:muiltiband }.

\end{appendix}


\end{document}